\documentclass[preprint,pre,aps]{revtex4}
\usepackage[dvips]{graphicx}
\bibstyle{prsty}
\begin{document}
 \title{Self-assembly in mixtures with competing interactions }
 \author{O. Patsahan}
 \affiliation{Institute of Condensed Matter Physics, Lviv, Ukraine}
\author{M. Litniewski and  A. Ciach}
\affiliation{Institute of Physical Chemistry,Polish Academy of Sciences, 01-224 Warszawa, Poland}
 \date{\today} 
 \begin{abstract}
 A binary mixture of particles interacting with spherically-symmetric potentials leading to microsegregation is studied by theory and molecular dynamics (MD) simulations. We consider spherical particles with equal diameters and volume fractions. Motivated by the mixture of oppositely charged particles with different adsorption preferences immersed in near-critical binary solvent, we assume short-range attraction long-range repulsion for the interaction between like particles, and short range repulsion long-range attraction for the interaction between different ones. In order to predict structural and thermodynamic properties of such complex mixtures, we develop a theory combining the density functional and field-theoretic methods. We show that concentration fluctuations in mesoscopic regions lead to a qualitative change of the phase diagram compared to mean-field predictions. Both theory and MD simulations show coexistence of a low-density disordered phase with a high-density phase with alternating layers rich in the first and the second component. The density and the degree of order of the ordered phase decrease with increasing temperature, up to a temperature where the theory predicts a narrow two-phase region with increasing density of both phases for increasing temperature. MD simulations show that  monocrystals of the solid and liquid crystals have  a prolate shape with the axis parallel  to the direction of concentration oscillations, and the deviation from the spherical shape increases with increasing periodic order.
 \end{abstract}
 \maketitle
  \section{Introduction}
  Competing interactions between particles in suspensions may lead to variety of patterns formed either by individual particles, or by assemblies with well defined size and shape. In particular, core-shell particles adsorbed at fluid interfaces form different highly ordered patterns for different thickness and structure of the polymeric shell, and different area fractions~\cite{rey:16:0,rauh:17:0,grishina:20:0}. In this case, effective repulsion induced by overlapping soft shells competes with effective attraction induced at larger distances by capillary forces. On the other hand, charged particles in solvents containing depletion agents attract or repel each other at short or at large distances, respectively~\cite{stradner:04:0,Bartlett2005}.
  This short-range attraction long-range repulsion (SALR) interaction is known also as the 'mermaid' potential due to the attractive head and the repulsive tail~\cite{royall:18:0}.  When the volume fraction of the particles increases, spherical or cylindrical clusters, or slabs are formed, and these assemblies can be periodically distributed in space at sufficiently low temperature ~\cite{ciach:08:1,ciach:10:1,ciach:13:0,zhuang:16:0,zhuang:16:1,edelmann:16:0,pini:17:0}. When the volume fraction further increases, voids instead of  the assemblies are formed in a reverse order~\cite{lindquist:16:0}. Notably, the sequence of the ordered phases in the SALR systems is universal, i.e. it is independent of the details of the interacting potential, and is the same as in amphiphilic systems~\cite{ciach:13:0}. Note that outside the hard cores,  the interaction between the core-shell particles is a 'negative' of the SALR (or the mermaid) potential, and can be memorized as a 'peacock potential' because of the repulsive head and  the attractive tail. 
  
  The ordered phases in the SALR system were predicted theoretically~\cite{ciach:08:1,archer:08:0,edelmann:16:0,pini:17:0} and observed in simulations \cite{candia:06:0,zhuang:16:0,zhuang:16:1,lindquist:16:0}.
 The rich variety of long-lived metastable states make it difficult to reach the equilibrium, however, and in simulations it was necessary to use intelligent tricks to obtain the stable phases~\cite{zhuang:16:0,zhuang:16:1}. Similar difficulties are present in experiments, and the ordered phases have not been detected yet~\cite{royall:18:0}. A very special methodology is necessary for obtaining the true equilibrium. For this reason, various systems and scenarios were suggested in order to overcome the problem of long-lived metastable states in experiment. In Ref.\cite{marlot:19:0,marlot:20:0} it was suggested to investigate colloid particles with surfaces preferentially adsorbing one component of a near-critical mixture in which the particles were suspended. Critical concentration fluctuations in the mixture confined between selective surfaces induce the so called thermodynamic Casimir potential between these surfaces. The potential is attractive for surfaces with like- and repulsive between surfaces with different  adsorption preferences~\cite{hertlein:08:0,gambassi:09:0}. 
 Charged particles interact in addition with screened Coulombic forces that are repulsive for like charges. The screened electrostatic interactions compete in this system with the thermodynamic Casimir potential induced by the critical fluctuations in the solvent. The sum of the Casimir and the electrostatic interactions between like particles has the SALR form when  the charge is relatively small, and the screening length is larger than the decay length of the Casimir potential~\cite{ciach:16:0}. The latter can be finely tuned by temperature. Because the repulsion can be easily turned on and off at a controlled speed, this system may be a good candidate for experimental detection of the ordered phases in the SALR systems. 
 
A suspension of  charged selective particles in a near-critical mixture offers many other opportunities for spontaneous formation of ordered patterns. Let us consider a binary mixture of colloid particles suspended in such a complex solvent. Let the particles of the first kind  adsorb one component of the binary solvent, and the particles of the second kind adsorb the other component, and let the different particles be oppositely charged. When the charge and temperature are tuned so that the like particles interact with the mermaid potential, then the interaction between different particles has the form of the peacock potential, 
because the Casimir force dominating at shorter distances is repulsive, and the electrostatic force dominating at larger distances is attractive.  As both types of interactions lead to formation of various  patterns, one can expect  interesting structural and thermodynamic properties in the system where both types of interactions are present.
 To memorize the complex interactions in such systems more easily, we may imagine 'two mermaids and a peacock' (2MP).

Motivated by the properties of the above binary mixture of particles in the binary critical solvent, we focus in this work on a general case of the mixture with the mermaid potential between like- and the peacock potential between different particles, since the 2MP potentials may be present in different soft-matter systems as well. We intend to determine these structural and thermodynamic features of such systems that are not limited to any specific shape of the interactions, but are common to all 2MP systems. As the SALR systems have a universal topology of the phase diagram, but the patterns in the core-shell particles depend sensitively on the shape of the interaction potential, the question to what extent the properties of the 2MP systems are universal or specific is open. 

A mixture of the above kind was studied in Ref.\cite{ciach:20:1} for a particular choice of the interaction potentials by both theory and   Monte Carlo simulations. Chains of alternating clusters composed of particles of the first kind followed by the clusters composed of the particles of the second kind  were observed at relatively high temperature, where the  long-range order was absent.
 Theoretical predictions for the correlation functions  were obtained  within the mesoscopic approach~\cite{ciach:08:1,ciach:11:2} combining the density functional and the  Brazovskii-type  field theories~\cite{evans:79:0,brazovskii:75:0}. The theoretical results agreed with the Monte Carlo simulations on a semi-quantitative level. The phase diagram, however, has not been determined yet. It is a purpose of this work.

It is important to stress that in the case of systems with spontaneous inhomogeneities the predictions of the mean-field (MF) theories are not correct at high temperatures where fluctuations play an important role. Even the sequence of the phases in the SALR systems is not correctly predicted on the MF level, except at low T. The continuous transition between the disordered and ordered phases predicted in MF turns out to be fluctuation-induced first order~\cite{brazovskii:75:0}.
 For this reason it is necessary to take the fluctuations into account, by which the theory becomes more complex~\cite{ciach:18:0}. 

In this work we further develop the mesoscopic theory for inhomogeneous mixtures~\cite{ciach:11:2}. We make assumptions based on physical grounds that allow to calculate the phase diagram for a symmetrical mixture of particles of identical size and of equal volume fractions with reasonable effort. 
We calculate the correlation functions, and determine the phase diagram for a particular choice of the interactions of the particles both in MF and beyond it. The theoretical predictions are compared with molecular dynamics (MD) simulations. The simulations, however, are performed for particles that are not composed of hard cores, but rather interact with the repulsive part of the Lenard-Jones potential. 
At large separations, the shapes of the  interaction potentials in simulations and theory are very similar, but are not the same at short separations. On the one hand,  we choose the potentials convenient for the theoretical and the simulation studies. On the other hand, from our approximate theory it follows that the phase diagram should depend only on some gross features of the interactions, when appropriate units for  temperature are used.
By comparison with the MD simulations for somewhat different shape of the potential, we will verify if the predictions of the theory with the Brazovskii-type approximation for the effects of fluctuations are valid.

The interaction potentials used in theoretical calculations and in simulations are presented and characterized in sec.~\ref{model}. The theoretical and simulation methods 
 are presented in sec.~\ref{theory} and sec.\ref{simulations}, respectively.
 In sec.~\ref{theory}a, we present the general formalism of our theory. In sec. ~\ref{theory}b and ~\ref{theory}c,  we derive the equations for the correlation functions, the chemical potential and the grand thermodynamic potential in MF and in the Brazovskii-type approximation, respectively.
 Readers not interested in the formalism, may skip these sections.
Sec.~\ref{results} contains our results, and in the last section our results are summarized and discussed.

 \section{ the model}
 \label{model}
 We consider a binary mixture of spherical particles with the particle diameter $a$, and assume that the interaction potential  $u_{ij}$  with $i,j=1,2$
 consists of strong repulsion for   the center-to-center distance $r\le a$, and of competing interactions for $r>a$.  In the following, we consider dimensionless distance $r^*=r/a$, and omit the asterisk for clarity. We assume the same interaction between like particles of both species, $u_{ii}(r)=u(r)$ that for $r>1$ has the form of  the short-range attraction and long-range repulsion.  For the interaction between different particles, we assume for $r> 1$ short-range repulsion and long-range attraction.
 
In the theory, we assume hard cores of the particles and $u_{ij}(r)= \infty$  for $r<1$. For $r>1$, we assume $u_{12}(r)=-u(r)$, with
  \begin{equation} 
  \label{V(r)a}
 u(r)=
-\frac{6\epsilon}{r^6}+A_Y\frac{e^{-r/2}}{r}.
\end{equation}
 
\begin{figure}
\includegraphics[scale=0.35]{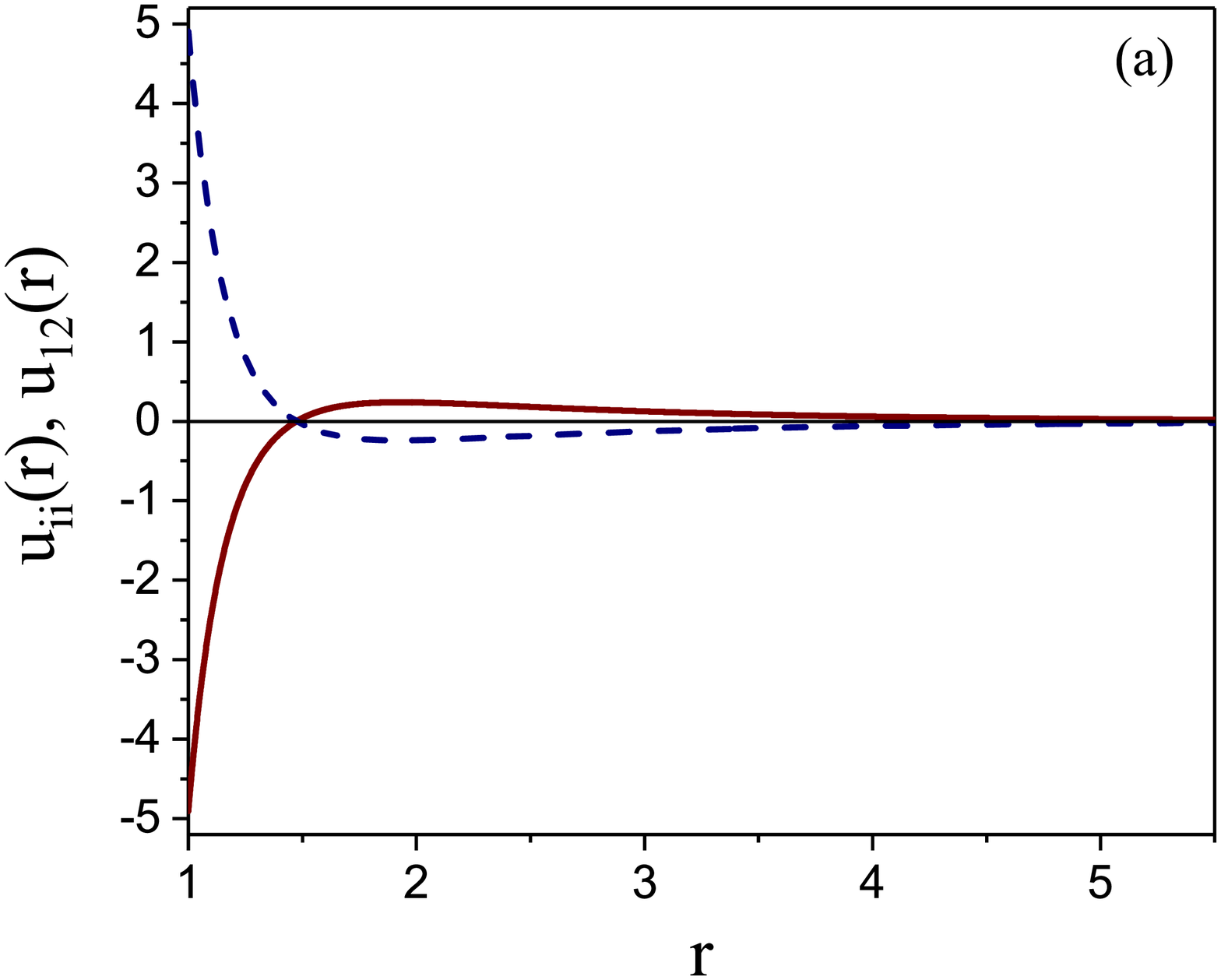}
\includegraphics[scale=0.35]{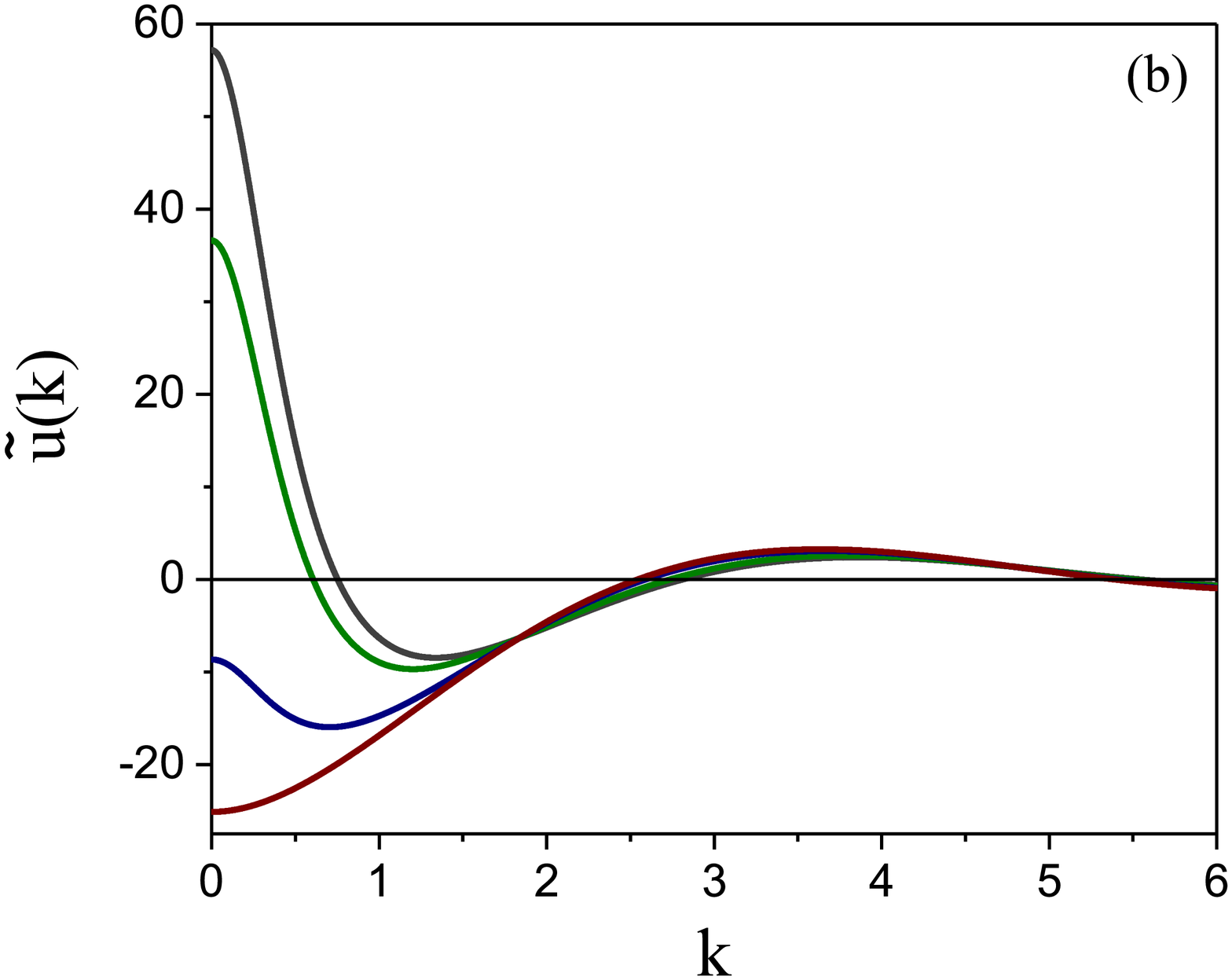}
\caption{(a) the interaction potential $u_{ij}$,  Eq.~(\ref{V(r)a}),  between like  particles (solid line) and between different particles (dashed line) for $A_Y=1.8$; (b) the interaction potential $\tilde u(k)$ in Fourier representation. 
From the bottom to the top line  $A_Y=0,0.36, 1.35, 1.8$, respectively. $u$, $r$ and $k$ are  in units of $\epsilon$, the particle diameter $a$ and $a^{-1}$, respectively.
}
\label{figVk}
\end{figure}

\begin{figure}
\includegraphics[scale=0.35]{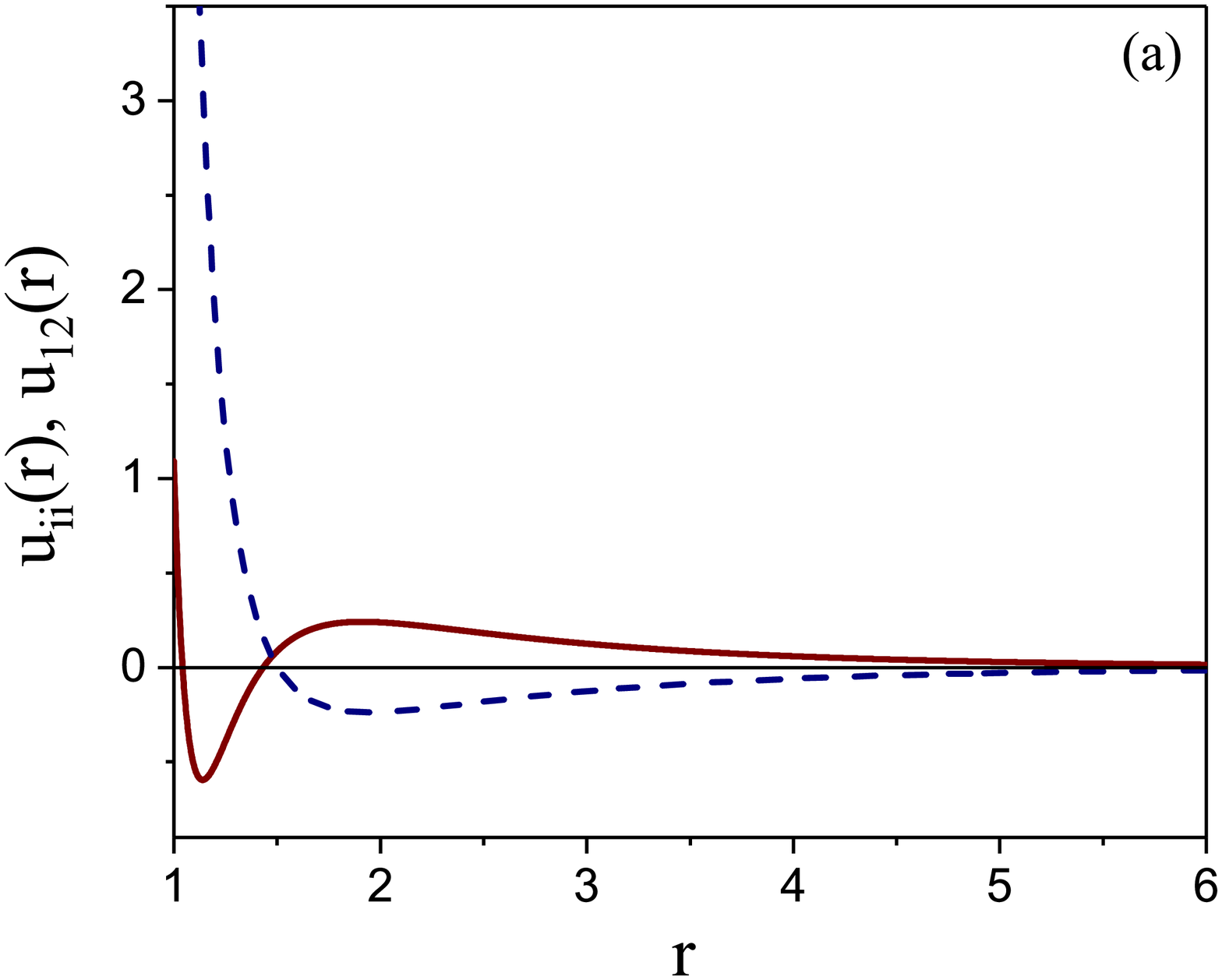}
\includegraphics[scale=0.35]{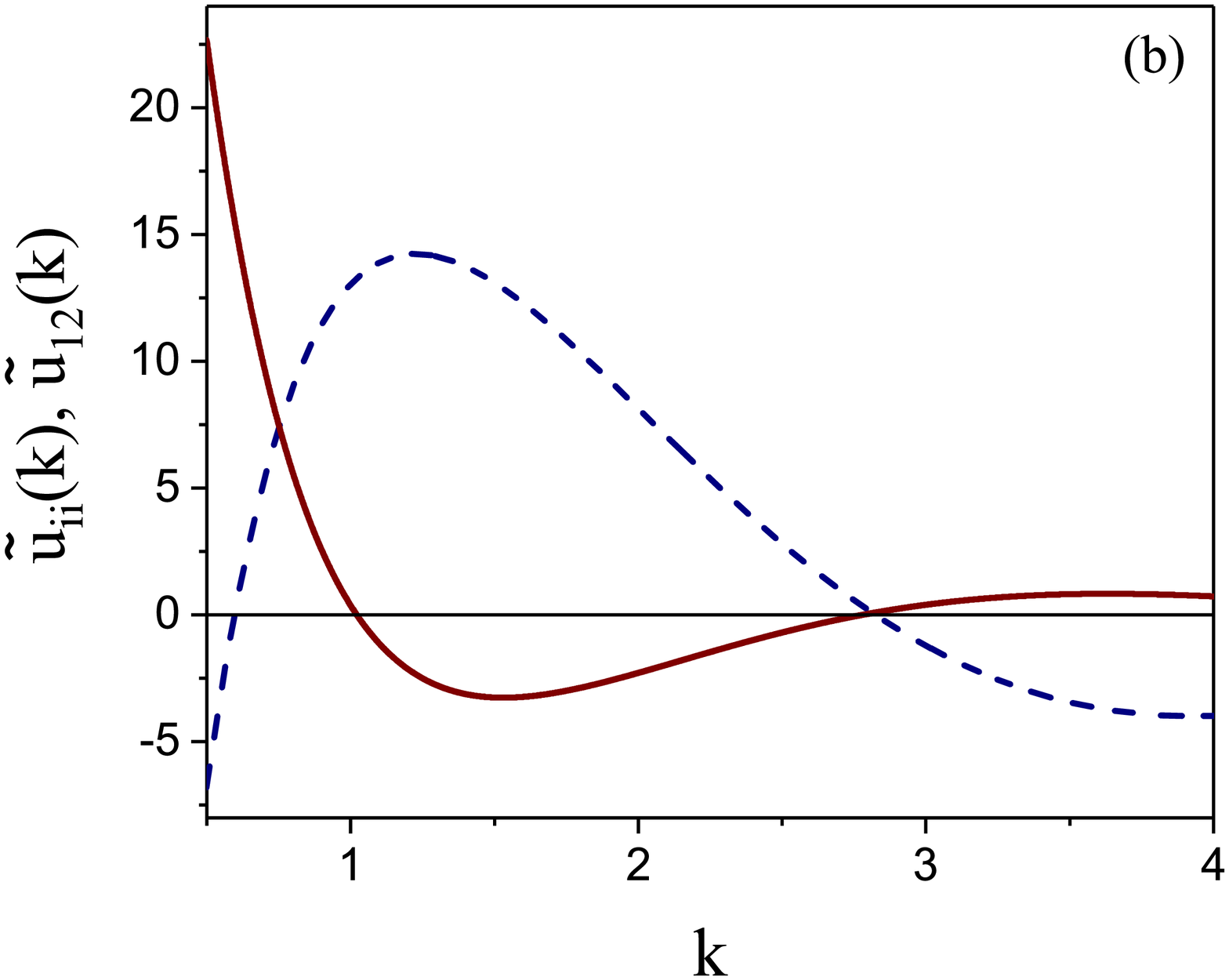}
\caption{the interaction potential between like  particles (solid line) and between different particles (dashed line) used in the MD simulations  (see Eqs.(\ref{uMD})-(\ref{uMD12}) for $A_Y=1.8$) in real space (a) and in Fourier representation (b). $u$, $r$ and $k$  are  in units of $\epsilon$  the particle diameter $a$,  and $a^{-1}$, respectively.}
\label{figVkMD}
\end{figure}

In the MD simulations, we  assume 
\begin{equation}
\label{uMD}
u_{ii}^{MD}=\frac{6\epsilon}{r^{12}}-\frac{6\epsilon}{r^6}+ A_Y\frac{e^{-r/2}}{r}
\end{equation}
and 
\begin{equation}
\label{uMD12}
u_{12}^{MD}=\frac{6\epsilon}{r^{12}}+ \frac{6\epsilon}{r^6}-A_Y\frac{e^{-r/2}}{r}.
\end{equation}
The hard core at $r=1$ present in the theoretical model, is replaced by the strong repulsion, $6\epsilon/ r^{12}$, and the size of the particle core is not uniquely defined. Because the short-range repulsion is significantly softer than the hard core, the shapes of the potentials in the theory and in the simulations for $r<1.5$  are different, as shown in Figs.~\ref{figVkMD}a and \ref{figVk}a. For $A_Y=1.8$ the minimum of $u_{ii}^{MD}(r)$ occurs at $r_{min}\approx 1.14$, and $|u_{ii}^{MD}(r_{min})|\ll |u(1)| $. At large distances, however, $u_{ii}^{MD}(r)\approx -u_{12}^{MD}(r)$ as assumed in our theory, and  $u_{ii}^{MD}(r)\approx  u(r)$. 
 
The strength of the repulsion $A_Y$ can be varied to model varying charge of the particles.  
For $A_Y=0$, the potential $u_{ij}$ reduces to attractive interactions between like particles, and to a repulsion between different particles, leading to a macroscopic phase separation. On the other hand, for  $\epsilon=0$ the potential $u(r)$ (Eq.(\ref{V(r)a})) reduces to the screened Coulomb potential between charged colloid particles.

The potential outside the hard core, $u(r)\theta(r-1)$, in Fourier representation is denoted by $\tilde u(k)$,  and takes the form
 shown in Fig.\ref{figVk}b for a few values of $A_Y\le 1.8$.
Note that in the absence of the repulsion, the potential takes the  minimum for $k=0$, whereas for sufficiently strong repulsion, the minimum occurs for $k_0>0$ that increases with increasing $A_Y$.
 
  $\tilde u(k)$
  describes the increase of the energy of the homogeneous  system when the concentration wave with the wavenumber $k$ is excited. The most probable concentration wave corresponds to the minimum of  $\tilde u(k)$; the most probable
  distance between like clusters or layers is $2\pi/k_0$, and the thickness of the aggregates is $\sim \pi/k_0$. Moreover, $\tilde u(k_0)$ sets the energy scale connected with the process of self-assembly. 
  
  To compare the relevant energy units in  the theory and  simulations, we should know $u_{ij}^{MD}$ outside the particle core in Fourier representation. Unfortunately, the size of the core is not uniquely defined. To have some insight, we Fourier transform $u_{ij}^{MD}(r)\theta(r-1)$ for $u_{ij}^{MD}$ given in Eqs.~(\ref{uMD})-(\ref{uMD12}). The results are shown in Fig.~\ref{figVkMD}b. Different forms of $u_{ii}^{MD}(r)$ and $-u_{12}^{MD}(r)$ at short distances lead to different positions and magnitudes of the minima of  $\tilde u_{ii}^{MD}(k)$ and $-\tilde u_{12}^{MD}(k)$ that are also somewhat different from the corresponding values in the theoretical model. 
  Moreover, the form of $\tilde u_{ij}^{MD}(k)$ depends strongly on the arbitrary choice of the particle diameter. Because of that, the energy scale for the process of self-assembly for softened particle cores is not uniquely defined. In addition, different forms of the potentials for $r<1.5$ lead to about ten times larger thermal energy $kT$ that is equal to the minimum of $u_{ii}(r)$, compared to the thermal energy equal to the minimum of $u_{ii}^{MD}(r)$ (see Figs.~\ref{figVk}a and \ref{figVkMD}a). 
  Thus, we will not try to compare the theoretical and simulation results on the quantitative level. 
  
  We choose for further calculations and simulations $A_Y=1.8$ which leads to small aggregates, $\pi/k_0\sim 2$. In this case, a relatively small number of particles gives a sufficiently large number of aggregates, and in turn leads to reliable results in simulations at a reasonable computational cost.

\section{ theory}
\label{theory}
\subsection{the formalism for symmetrical mixtures}
\label{theoryA}

  In the mesoscopic approach developed for inhomogeneous mixtures in Ref.\cite{ciach:11:2}, we consider mesoscopic regions and mesoscopic states. In a particular mesoscopic state, the volume fraction of particles of the $i$-th species  around the point ${\bf r}$,  $\zeta_i({\bf r})$, is the fraction of the volume of the mesoscopic region occupied by the particles. The mesoscopic regions are comparable with or larger than $1$ (in $a$-units), and smaller than the scale of the inhomogeneities.  If we assume that 
  the dimensionless ``mass'' of the particle is homogeneously distributed over its volume $\pi/6$, then  $\zeta_i({\bf r})$  is a continuous function 
  of ${\bf r}$~\cite{ciach:18:0}. The functions representing the    local  
  concentration and  volume fraction, $c({\bf r})=\zeta_1({\bf r})-\zeta_2({\bf r})$ and
   $\zeta({\bf r})=\zeta_1({\bf r})+\zeta_2({\bf r})$ respectively,
   can be considered as constraints imposed on the microscopic states. 
    The concentration and the volume fraction  averaged over the system volume $V$ are denoted by $\bar c$ and  $\bar\zeta$, respectively.
   We limit ourselves to a symmetrical case, with   $\bar c=0$, and  consider a range of  $\bar\zeta$. 
   
 In theoretical considerations, it is  convenient to consider an open system, with fixed chemical potentials $\mu_1$ and $\mu_2$. In the symmetrical case, $\mu_1=\mu_2=\mu$.  We assume that   in the presence of the above constraints the grand potential 
   can be written as 
   \begin{equation}
    \label{Omco}
    \Omega_{co}[c,\zeta]=
   U_{co}[c,\zeta]-TS[c,\zeta] -\mu\int d{\bf r}\zeta({\bf r}),
   \end{equation}
where $T$ is temperature, and $S[c,\zeta]$ 
is the entropy.
We make the approximation
$-TS=\int d{\bf r} f_h(c({\bf r}),\zeta({\bf r}))$, where $f_h(c,\zeta)$ is the free-energy density 
of the hard-core reference system in the local-density
approximation,
\begin{equation}
\beta f_h(c,\zeta)=\zeta_1\ln\zeta_1+\zeta_2\ln\zeta_2+\beta f_{ex}(\zeta).
\end{equation}
 In the particular case of the Carnahan-Starling approximation,
\begin{equation}
\beta f_{ex}(\zeta)=
\rho\Big[
\frac{4\zeta-3\zeta^2}{(1-\zeta)^2}-1
\Big],
\end{equation}
where $\rho=6\zeta/\pi$.
Finally,
 \begin{equation}
    \label{Uco}
    U_{co}[c,\zeta]=
    \frac{1}{2}\int d{\bf r}_1\int d{\bf r}_2\zeta_i({\bf r}_1)V_{ij}(|{\bf r}_1-{\bf r}_2|)\zeta_j({\bf r}_2)=\frac{1}{2}\int d{\bf r}_1\int d{\bf r}_2c({\bf r}_1)V(|{\bf r}_1-{\bf r}_2|)c({\bf r}_2)
   \end{equation}
is the internal energy for the assumed type of  interactions, and summation convention for repeated indexes is used.  Because $\zeta_i=\pi\rho_i/6$ is used in the above definition, we have rescaled the interaction potential, $V_{ij}=u_{ij}(6/\pi)^2$. 

 In Fourier representation,
\begin{equation}
 \label{UcoF}
    U_{co}[c,\zeta]= \frac{1}{2}\int \frac{d{\bf k}}{(2\pi)^3}\tilde c({\bf k})\tilde V(k)\tilde c(-{\bf k}).
\end{equation}

Importantly, $\tilde V(k)=(6/\pi)^2\tilde u(k)$ takes the minimum for $k=k_0>0$ when $A_Y>0$. Note that the ordering effect of the energy concerns only the concentration, and the largest energy gain is for the concentration wave with the wavenumber $k_0$. 

When the constraints imposed on $c({\bf r})$ and
   $\zeta({\bf r})$ are released, the grand potential contains a fluctuation contribution and has the form~\cite{ciach:11:2}
   
\begin{eqnarray}
 \label{Ome0}
  \beta \Omega[c,\zeta]= \beta \Omega_{co}[c,\zeta]-\ln\Bigg[
  \int D\phi  \int D\psi e^{-\beta H_f[c,\zeta;\phi,\psi]}
  \Bigg],
 \end{eqnarray}
 where 
  $ H_f[c,\zeta;\phi,\psi]=\Omega_{co}[c+\phi,\zeta+\psi]-\Omega_{co}[c,\zeta]$ is  associated with appearance of the fluctuation $\phi$ of the local concentration, and the fluctuation $\psi$ of the local volume fraction. For equilibrium $c({\bf r})$ and
   $\zeta({\bf r})$, we have $\langle \phi\rangle=0=\langle \psi\rangle$, and $\beta \Omega[c,\zeta]$ takes the global minimum for fixed $T$ and $\mu$. The necessary condition for the minimum of $\Omega_{co}[c,\zeta]$ has the form
   \begin{eqnarray}
\label{mincond1}
 \frac{\delta \beta\Omega[c,\zeta]}{\delta \zeta({\bf r})}=\frac{\delta \beta\Omega_{co}[c,\zeta]}{\delta\zeta({\bf r})}
 +\Bigg\langle \frac{\delta \beta H_f}{\delta\zeta({\bf r})}\Bigg\rangle=0
\end{eqnarray}
and
\begin{eqnarray}
\label{mincond2}
 \frac{\delta \beta\Omega[c,\zeta]}{\delta c({\bf r})}=\frac{\delta \beta\Omega_{co}[c,\zeta]}{\delta c({\bf r})}
 +\Bigg\langle \frac{\delta \beta H_f}{\delta c({\bf r})}\Bigg\rangle=0.
\end{eqnarray}
 In the ordered phases, $c({\bf r})$ and
   $\zeta({\bf r})$ are periodic functions of ${\bf r}$.  The periodic phases can correspond either to solid  or to liquid crystals, depending on the degree of order.
  We shall use the term 'periodic phase' for any phase with periodic $c({\bf r})$.
   
     In the disordered fluid phase, both functions are position independent. For fixed $T$ and $\mu$, the two phases coexist when the grand potentials of these phases,  $\Omega=-pV$, where $p$ is pressure, are equal.
   
   The correlation functions $G_{ij}$ for $\zeta_i,\zeta_j$ are the matrix elements of ${\bf G}={\bf C}^{-1}$,
where the elements $C_{ij}$ of the matrix ${\bf C}$ are the second functional derivatives of $\beta\Omega[c,\zeta]$ with respect to $\zeta_i$ and $\zeta_j$. Because of the symmetry of the interactions,
the eigenvectors of the matrix ${\bf C}$ are the $c$ and $\zeta$ fields. In this case, we consider
$G_{cc}(r)=\langle c({\bf r}_1) c({\bf r}_2)\rangle$ and $G_{\zeta\zeta}(r)=\langle \zeta({\bf r}_1) \zeta({\bf r}_2)\rangle-\bar\zeta^2$. The above correlation functions  in Fourier representation are simply given by
$\tilde G_{cc}(k)=1/\tilde C_{cc}(k)$ and  $\tilde G_{\zeta\zeta}(k)=1/\tilde C_{\zeta\zeta}(k)$.

   \subsection{Mean field approximation}
   \label{theoryB}
    In this subsection we limit ourselves to the MF approximation, where the last term in Eq.~(\ref{Ome0}) is disregarded. The correlation functions in the disordered phase are simply given by 
\begin{equation}
\label{GMF}
\tilde G_{cc}^{MF}(k)^{-1}=\tilde C_{cc}^{MF}(k)=\beta \tilde V(k)+1/\bar\zeta,
\end{equation}
\begin{equation}
\tilde G_{\zeta\zeta}^{MF}(k)^{-1}=\tilde C_{\zeta\zeta}^{MF}(k)=1/\bar\zeta+\partial^2 f_{ex}(\bar\zeta)/\partial \bar\zeta^2.
\end{equation}
Note that in this MF approximation, $\tilde C_{\zeta\zeta}^{MF}(k)$ is independent of $k$. This means strictly local correlations. 

In MF, the disordered fluid looses stability with respect to a periodic $c({\bf r})$ with the wavenumber $k_0$ along the so called $\lambda$-line given by $\tilde C_{cc}^{MF}(k_0)=0$, i.e.
\begin{equation}
\bar T_{\lambda}=-\tilde V(k_0)\bar\zeta, 
\end{equation}
where in the case of competing interactions, $\tilde V(k_0)<0$. This instability can be preempted by a first-order transition to an ordered phase with periodically distributed particles (colloidal crystal or liquid crystal). 

Let us first consider the disordered phase, where $c=0$, and $\zeta=\bar\zeta$ is determined by the minimum of 
$\Omega_{co}[0,\zeta]$. Minimization of Eq.(\ref{Omco}) with respect to $\zeta$ gives
\begin{equation}
\label{Omg}
\beta\Omega_g/V=\beta f_{ex}(\bar\zeta)-A_1(\bar\zeta)\bar\zeta-\bar\zeta,
\end{equation}
where $\Omega_g$ denotes the grand potential in the disordered phase and $\bar\zeta$ is the solution of the equation
\begin{equation}
\label{bmuMF}
\beta\mu=\ln\Big(\frac{\bar\zeta}{2}\Big) +1+ A_1(\bar\zeta).
\end{equation}
Here and below,  
\begin{equation}
A_n(\zeta)=\frac{d^n \beta f_{ex}(\zeta)}{d\zeta^n}.
\end{equation}
Note that in the symmetrical mixture with the internal energy depending only on the concentration, $\beta\mu$ is independent of temperature in this MF approximation.

In the case of the periodic phase, we postulate that in the symmetrical case, the concentration is a periodic function with oscillations in one direction, say $z$. We assume that 
\begin{equation}
c(z)=\Phi g_c(z),\hskip1cm \zeta(z)=\bar\zeta+\Psi g_{\zeta}(z),
\end{equation}
 with $\bar\zeta=\int_0^P\zeta(z)dz/P$,  $\int_0^P g(z)dz/P=0$ and $ \int_0^P g(z)^2dz/P=1$ for $g=g_c,g_{\zeta}$, where $P$ denotes the period of oscillations of $c$. The period of $g_{\zeta}$ is $P/2$
 because of the symmetry of the model. The oscillations of $\zeta$ appear because of the coupling between $c$ and $\zeta$ in the entropy of mixing.

The problem simplifies greatly, if we restrict ourselves to relatively high $ T$, where $\Phi$ is small, and we can make the assumptions
$g_c(z)=\sqrt 2 \cos(k_0 z)$ and $g_{\zeta}(z)=\sqrt 2 \cos(2k_0 z)$. Such sinusoidal shapes were indeed observed in one-component SALR systems for not very low $ T$~\cite{pini:17:0}. With the above assumption, we have to minimize a function of 3 variables, $\Phi, \Psi$ and $\bar\zeta$. We have:
\begin{equation}
\label{Omegar}
\frac{1}{V}\frac{\partial \beta \Omega_c}{\partial \bar\zeta}=\frac{1}{P}\int_0^P \frac{\partial \beta f_h}{\partial \zeta(z)}dz-\beta \mu=0,
\end{equation}

\begin{equation}
\label{OmegaPsi}
\frac{1}{V}\frac{\partial \beta \Omega_c}{\partial \Psi}=\frac{1}{P}\int_0^P \frac{\partial \beta f_h}{\partial \zeta(z)}g_{\zeta}(z)dz=0,
\end{equation}
\begin{equation}
\label{OmegaPhi}
\frac{1}{V}\frac{\partial \beta \Omega_c}{\partial \Phi}=
\tilde V(k_0)\Phi+
\frac{1}{P}\int_0^P \frac{\partial \beta f_h}{\partial c(z)}g_c(z)dz=0,
\end{equation}
where $\Omega_c$ is the grand potential in the ordered (periodic) phase.
We Taylor-expand $\beta\Omega_c$ in terms of $\Phi$ and $\Psi$. From (\ref{OmegaPsi}), we obtain the relation between $\Psi$ and $\Phi$,
\begin{equation}
\label{Psi}
\Psi= \frac{\sqrt 2 \Phi^2}{4\bar\zeta(1+\bar\zeta A_2(\bar\zeta))}+O(\Phi^4).
\end{equation}
From (\ref{OmegaPhi}) and (\ref{Psi}) we obtain 
\begin{equation}
\label{Phi}
\Phi^2=-\frac{4\bar\zeta^2(1+\beta\tilde V(k_0)\bar\zeta)(1+\bar\zeta A_2(\bar\zeta))}{1+2\bar\zeta A_2(\bar\zeta)}.
\end{equation}
The solution of (\ref{Phi}) is meaningful for $1+\beta\tilde V(k_0)\bar\zeta<0$, i.e. in the region where the disordered fluid is unstable.
From (\ref{Omegar}) we obtain, 
keeping terms up to $O(\Phi^4)$, 
\begin{equation}
\label{mu}
\beta\mu\approx \ln\Big(\frac{\bar\zeta}{2}\Big)+1+A_1(\bar\zeta)-\frac{1}{2\bar\zeta^2}\Phi^2+\frac{1}{2}\Bigg(A_3(\bar\zeta)-\frac{1}{\bar\zeta^2}\Bigg)\Psi^2+\frac{\sqrt 2}{2\bar\zeta^3}\Psi\Phi^2-\frac{3}{8\bar\zeta^4}\Phi^4.
\end{equation}
Because $\Phi$ depends on $\beta\tilde V(k_0)$, $\beta\mu$ is temperature dependent  in the periodic phase in MF.
Finally, the grand potential in the periodic phase takes the form
\begin{equation}
\label{Omaprox}
\beta\Omega_c\approx \bar\zeta\ln\Big(\frac{\bar\zeta}{2}\Big) +\beta f_{ex}(\bar\zeta)-\beta\mu\bar\zeta +\frac{1}{2}\Bigg(\beta\tilde V(k_0)+\frac{1}{\bar\zeta}\Bigg)\Phi^2
+\frac{1}{2}\Bigg(\frac{1}{\bar\zeta}+A_2(\bar\zeta)\Bigg)\Psi^2-\frac{\sqrt 2}{4\bar\zeta^2}\Psi\Phi^2 +\frac{1}{8\bar\zeta^3}\Phi^4 +O(\Phi^6).
\end{equation}

From the above equation it follows that the natural variables are $\beta\mu$ and $\beta\tilde V(k_0)$. Thus, we will consider  $\beta\mu$ and  $T^*=kT/|\tilde V(k_0)|$. Note that in this MF approximation, the grand potential depends on the interaction potential only through $\tilde V(k_0)$. This means universal phase diagrams with properly scaled temperature.
 
By inserting  $\Psi$ and $\Phi$ given by (\ref{Psi}) and (\ref{Phi}) in (\ref{mu}) and (\ref{Omaprox}), and by eliminating $\bar\zeta$ from (\ref{Omaprox}) and (\ref{mu}), we obtain  $\beta\Omega_c$ as a function of $T^*$ and $\beta\mu$. The stable phase for given $T^*$ and $\beta\mu$ is the one corresponding to the smaller value of the grand potential. We obtain the MF phase diagram by comparing $\beta\Omega_c$ with $\beta\Omega_g$ for fixed $T^*$ and $\beta\mu$ in sec.\ref{results} 

\subsection{The Brazovskii-type theory for symmetrical mixtures}
\label{theoryC}
The MF is not expected to give  correct results for high $T$, where the fluctuations play an important role and lead to formation of delocalized aggregates. Thus, the more accurate expression for $\Omega$, Eq.(\ref{Ome0}), should be considered. Because the energy gain concerns the local deviations of the concentration from $\bar c$, and the  deviations of $\zeta$ from $\bar\zeta$ do not directly influence the energy, we expect that fluctuations of the former are much more probable and of larger magnitude than fluctuations of the latter. 
 If the fluctuations of the local concentration are of a significantly larger magnitude than the fluctuations of the local volume fraction, we can consider a simplified theory, where the fluctuations of the local concentration are taken into account, but the fluctuations of the total volume fraction are disregarded. 
  
In the approximation with only the concentration fluctuations taken into account, the grand potential (\ref{Ome0}) takes the form  
\begin{eqnarray}
 \label{Ome}
  \beta \Omega[c,\zeta]= \beta \Omega_{co}[c,\zeta]-\ln\Bigg[
  \int D\phi e^{-\beta H_f[c,\zeta;\phi]}
  \Bigg].
 \end{eqnarray}

  Following  Ref.\cite{ciach:18:0}, we make the approximation
 \begin{eqnarray}
 \label{Hf3}
\beta \bar H_f[c,\zeta;\phi]\approx
 \frac{1}{2}\int \frac{d{\bf k} }{(2\pi)^3}\tilde \phi({\bf k})\beta\tilde V(k)\tilde \phi(-{\bf k})+\int d{\bf r}\Bigg[\frac{A_{0,2}(\bar\zeta,\Phi)}{2}\phi({\bf r})^2+
\frac{a_{0,4}(\bar\zeta)}{4!}\phi({\bf r})^4
 \Bigg],
\end{eqnarray}
where
 \begin{eqnarray}
 \label{A02}
  A_{0,2}(\bar\zeta,\Phi)\approx a_{0,2}(\bar\zeta)+\frac{a_{0,4}(\bar\zeta)\Phi^2}{2}
 \end{eqnarray}
and 
\begin{equation}
 a_{m,n}(\zeta)=\frac{\partial^{n+m}(\beta f_h)}{\partial^n c\partial^m \zeta}|_{c=0}.
\end{equation}

In order to calculate the fluctuation contribution to $\Omega$, we make the approximation~\cite{ciach:12:0}
\begin{equation}
\label{gau}
e^{-\beta H_f}=e^{-\beta H_G}(1-\beta\Delta H),
\end{equation}
where $H_G$ has the  Gaussian form
\begin{equation}
\label{gauss}
\beta H_G=\frac{1}{2}\int \frac{d{\bf k} }{(2\pi)^3}\tilde \phi({\bf k})\tilde C_{cc}(k)\tilde \phi(-{\bf k}),
\end{equation}
and  we have assumed that $\Delta H=H_f-H_G$ is small.
For  $\Omega$ approximated by (\ref{Ome}), 
 $\tilde C_{cc}(k)$
 contains the fluctuation contribution.
Finally, we approximate $\langle X\rangle$ by averaging the quantity $X$  with the probability
$\propto \exp(-\beta H_G)$. 
 In order to calculate the fluctuation contributions in (\ref{Ome}), (\ref{mincond1}) and (\ref{mincond2}), we need to know $\tilde C_{cc}(k)$.
 In the  Brazovskii-type approximation, it obeys the equation~\cite{brazovskii:75:0,ciach:18:0,ciach:12:0}
\begin{equation}
\label{Ccc4}
 \tilde C_{cc}(k)=\beta \tilde V(k) +
 A_{0,2}(\bar\zeta,\Phi)+\frac{a_{0,4}(\bar\zeta)}{2}{\cal G} .
\end{equation}
The form of ${\cal G}=\langle\phi({\bf r})^2\rangle=(2\pi)^{-3}\int d{\bf k}\tilde G_{cc}(k)$ is well known for $\tilde V(k)$ assuming the minimum for $k=k_0>0$ from the previous studies \cite{ciach:12:0},
\begin{equation}
\label{calG}
{\cal G}=\frac{k_0^2}{\pi\sqrt{2\beta\tilde V^{''}(k_0)\tilde C_{cc}(k_0)}},
\end{equation}
and  $\tilde C_{cc}(k_0)$ is the solution of (\ref{calG}) and (\ref{Ccc4}) with $k=k_0$. The explicit expression for $\tilde C_{cc}(k_0)$ is given in Ref.\cite{ciach:12:0,ciach:18:0}.

Using (\ref{Ome}), (\ref{gau}), (\ref{calG}) and (\ref{gauss}) we obtain (see Ref.\cite{ciach:12:0} for more details)
\begin{eqnarray}
 \label{Omee}
  \beta \Omega[c,\zeta]/V= \beta \Omega_{co}[c,\zeta]/V +\tilde C_{cc}(k_0){\cal G}
  -\frac{a_{0,4}(\bar\zeta){\cal G}^2}{8}.
 \end{eqnarray}

We make the same assumptions concerning $c$ and $\zeta$ as in MF, and we need to minimize $\Omega$ with respect to $\bar\zeta$, $\Phi$ and $\Psi$.
In this approximation, the fluctuation contribution is independent of $\Psi$ and (\ref{Psi}) holds, but
the minimum with respect to $\Phi$ gives 
\begin{equation}
\label{Phif}
\Phi^2(1+2\bar\zeta A_2(\bar\zeta))+4(\bar\zeta^2+\beta\tilde V(k_0)\bar\zeta^3+{\cal G})(1+\bar\zeta A_2(\bar\zeta))=0.
\end{equation}
Note that ${\cal G}$ is a function of $\Phi$ (see (\ref{calG}), (\ref{Ccc4}) and (\ref{A02})).
In this Gaussian approximation,  we obtain from (\ref{mincond1}) and (\ref{Hf3}) the chemical potential
\begin{eqnarray}
\label{muf}
\beta\mu& \approx & \ln\Big(\frac{\bar\zeta}{2}\Big)+1+A_1(\bar\zeta)-\frac{1}{2\bar\zeta^2}\Phi^2+\frac{1}{2}\Big(A_3(\bar\zeta)-\frac{1}{\bar\zeta^2}\Big)\Psi^2+\frac{\sqrt 2}{2\bar\zeta^3}\Psi\Phi^2-\frac{3}{8\bar\zeta^4}\Phi^4
\nonumber
\\
& -&\Bigg(\frac{1}{2\bar\zeta^2}+\frac{3\Phi^2}{2\zeta^4}\Bigg){\cal G}-\frac{3}{4\bar\zeta^4}{\cal G}^2.
 \end{eqnarray}
 
 Eqs.(\ref{muf}), (\ref{Omee}), (\ref{calG}) and (\ref{Ccc4}) hold for both, the periodic and the disordered phase, with $\Phi=\Psi=0$ in the latter case. In contrast to the MF approximation, $\beta\mu$ depends on $T^*$ in the disordered phase (for $\Phi=\Psi=0$), because of the dependence on $T^*$ of $\cal G$.
 
 It is important to note that in contrast to the MF approximation, where the only dependence on the interaction potential is through $\tilde V(k_0)$, in this Brazovskii approximation, $\cal G$ and hence the phase diagram, depend in addition on $\tilde V^{''}(k_0)$.
 Still, just two parameters are sufficient to characterize the interactions in this theory. With  $\tilde V(k_0)$ setting the energy scale (i.e.  $T^*=kT/|\tilde V(k_0)|$), the same phase diagrams are expected for all interaction potentials with the same value of $\tilde V^{''}(k_0)/\tilde V(k_0)$.
 
  In order to obtain the phase diagram, we first calculate $\Psi$ and $\Phi$ from (\ref{Psi}) and (\ref{Phif}) respectively, and insert the results in (\ref{muf}) and (\ref{Omee}). From the last pair of equations, we obtain $\Omega$ as a function of $\mu$ and compare the solution for $\Phi=\Psi=0$ with the solution with $\Phi\ne 0$. The results are presented in sec.\ref{results}. 

  \section{ simulations}
  \label{simulations}
 The simulations were performed using classical constant energy and volume method \cite{allen:90:0}. We considered the systems of $N=N_1+N_2$ particles with $N_1=N_2$, enclosed in a rectangular box, where $L_x$, $L_y$, $L_z$ give the length of the edges. The temperature was always kept constant by scaling the particle velocities once for a given time interval. As in the previous paper \cite{litniewski:19:0}, the  potentials (\ref{uMD}) and (\ref{uMD12}) were truncated at $r=r_c=6.75$. 
 
 Different numbers of particles and  four different boundary conditions (BC) were chosen: (i) periodic boundary conditions along the $x$ and $y$ directions, with the walls at $z=0$ and $z=L_z$ interacting with the particles; (ii)  the wall at $z = 0$ attractive for the first, and repulsive for the second component, and all the remaining walls of the simulation box repulsive; (iii) a box with all walls repulsive; and (iv)  periodic boundary conditions in $x,y,z$ dimensions.  
 
The BC of the type (i), were chosen for low temperature, in order to pin-point  the crystal to the left wall. 
 The system of $N=32000$, $L_x = L_y = 200$, $L_z = 800$ for $\bar T=kT/\epsilon=0.16$ was simulated with this BC. 
 Attraction of the component $1$ and repulsion 
  of the component $2$   from the left-hand side wall was assumed, with the potentials
  \begin{eqnarray}
  V_{attr}(z)&=&\frac{2\epsilon}{z^{12}}-\frac{2\epsilon}{z^{6}}, \label{A1}\\
   V_{rep}(z)&=&\frac{\epsilon}{2z^{12}}, \label{A2}
  \end{eqnarray}
  respectively. 
  $z$ denotes the distance from the left wall. The right-hand side wall repulses both components according to Eq.(\ref{A2}), with $z$ replaced by $L_z-z$.
  
  The BC of the type (ii), were used for a determination of the volume fraction and the concentration profiles in the periodic phase adsorbed at the wall at $z=0$ for $\bar T=0.33$. $N=85184$ particles, attraction of the first component to the $z=0$ wall equal to $2 V_{attr}(z)$, and repulsion of the second component from this wall equal to $ V_{rep}(z)$ (see (\ref{A1}) and (\ref{A2})) were assumed. All the remaining walls were repulsive with the potential (\ref{A2}), with $z$ replaced by the distance from the corresponding wall.
  
 In order to determine the solid/liquid-gas coexistence curve, two series ($1$ and $2$) of simulations at different temperatures for the two systems: System~$1$ for Series~$1$, with the BC of the type (iii), and System~$2$ for Series~$2$, with the BC of type (iv), were carried out. Each series started at a very low temperature, with the crystal enclosed in the simulation box.
 
  System~$1$ was designed to measure the crystal parameters. The simulations were performed for $N=26566$ enclosed in the small volume of $L_x=45$, $L_y=107$, $L_z=35$, with reflective BC imposed for all borders via the $V_{rep}$ potential (\ref{A2}), with $z$ representing the distance from the wall for all $6$ walls of the simulation box. 
  
  System~$2$ was typical for the simulations of the  two-phase systems~\cite{Watanabe2012,Morris2002}. It consisted of $N=30812$ particles enclosed in the box of $L_x=28$, $L_y=44$, $L_z=200$ with periodic BC in all directions. To verify the results for high $\bar T$, Series~$2$ was extended by two additional simulations (Series~$2a$) for larger systems: $N=110184$,  $L_x=L_y=65.73$, $L_z =173.44$ at $\bar T=0.34$ and $N=116544$,  $L_x=L_y=65.73$, $L_z=172.11$ at $\bar T=0.35$.  
 
 The simulations for $\bar T=0.40$
 were performed for $N=85184$
  and four densities, $N/V=0.1$, $0.2$, $0.3$, $0.4$, with periodic BC (type (iv)). 
 
 The density of both, the gas and the liquid phase could be easily determined directly from the density profile from Series~$2$. Unfortunately, the crystal usually does not adapt to the box shape and the method may not work correctly. For this reason the crystal density was determined by counting the mean number of particles placed in the spheres which center coincides with the center of the crystal mass. Six spheres with radii $R_{\mu}=5$, $6$, $7$, $8$, $9$, $10$ were considered. Using this method, one assumes that the density profile in the vicinity of the mass center is flat in the range larger than $R_{\mu}$. This was fulfilled for all $R_{\mu}$ up to $\bar T=0.32$. A very low decrease in the calculated density with increasing $R_{\mu}$ appeared only for $\bar T>0.32$; the effect, however, was very weak. The largest difference in the density for $R_{\mu}=8$ and $10$ appeared for $\bar T=0.35$, but it did not exceed $1\%$ of relative value. The method has been applied for all the state points from Series~$1$ and for $\bar T<0.27$ from Series~$2$. For the remaining state points, the densities were determined from the density profile form Series~$2$.
 
 The dense phase-gas equilibrium volume fractions from Series~$1$ and $2$ are presented in Table~\ref{Table}. The comparison of the densities shows significant inconsistencies between the results from Series~$1$ and $2$ for $\bar T=0.31$, $0.34$, $0.35$. The reasons for the inconsistencies are discussed below.
 \begin{table}[h]
 	\centering
 	\caption{\label{Table} The volume fraction of the dense phase, $\zeta_{dens}$, and the gas, $\zeta_{gas}$, as a function of the temperature $\bar T$  for Series~$1$ and $2$. All values of $\zeta_{dens}$ for  Series~$1$ and that for    $\bar T<0.27$ for Series~$2$ are obtained for $R_{\mu}=10$. The remaining values are obtained from density profiles. The bar means that the value is not measured. $\bar T=kT/\epsilon$ and  $\zeta=\pi\rho/6$, $\rho=N/V$}	
 	\begin{tabular}{ c  c  c  c  c  c } 
 		\hline
 		&Series~$1$ & \multicolumn{2}{c}{Series~$2$}&\multicolumn{2}{c}{Series~$2a$}  \\ \hline 
 		$\bar T$ & $\zeta_{dens}$ & $\zeta_{dens}$ & $\zeta_{gas}$ & $\zeta_{dens}$ &  $\zeta_{gas}$\\ \hline   
 		0.140 & 0.383 & - & -  & - & - \\ 
 		0.180 & 0.379 & - & - & - & - \\ 
 		0.220 & 0.374 & - &  - & - & - \\
 		0.250 & 0.370 & 0.370& 0.0014 & - & - \\
 		0.260 & 0.369 & 0.369& 0.0023 & - & - \\
 		0.270 & 0.281 & 0.278& 0.0034 & - & - \\
 		0.280 & 0.273 & 0.270& 0.0048 & - & - \\
 		0.290 & 0.263 & 0.259& 0.0067 & - & - \\
 		0.300 & 0.250 & 0.245& 0.0091 & - & - \\
 		0.310 & 0.233 & 0.208& 0.0120 & - & - \\
 		0.315 & 0.206 & 0.200& 0.0136 & - & - \\
 		0.320 & 0.197 & 0.192& 0.0160 & - & - \\
 		0.330 & 0.181 & 0.173& 0.0212 & - & - \\
 		0.340 & 0.165 & 0.152& 0.0285 & 0.152 & 0.0288 \\
 		0.350 & 0.152 & 0.128& 0.0414 & 0.127 & 0.0429 \\  
 		\hline		
 	\end{tabular}
 	\centering
 	\label{table}
 \end{table}

 The inconsistencies for $\bar T=0.34$ and $0.35$ come from the disadvantage of System~$1$ at high $\bar T$. For  System~$1$, the walls can be close to the crystal border, 
  which may influence the simulation results. We investigated the effect for $\bar T=0.27$, $0.31$, $0.32$, $0.35$ by performing additional simulations (Test) with the box enlarged to $L_x=55$, $L_y=112$, $L_z=45$. For $\bar T\le 0.32$, the increase of the system volume (by a factor of over $1.6$) influenced neither the density nor the structure of the dense phase. Also the shape and the volume occupied by the dense phase 
  remained nearly unchanged. For $\bar T=0.35$, however, consequences of the increase of the system volume were very significant: the measured density decreased by around $0.1$ relative value. This shows that for high $\bar T$ the volume of System~$1$ is too small, and the results are burdened with very high errors. In contrast to Series~$1$, the results from Series~$2a$ are in good agreement with that from Series~$2$, which strongly validates the latter.    
 
 The origin of the inconsistency for $\bar T=0.31$ is different.
  We found that the change of the temperature from $\bar T=0.31$ to $0.315$ for Series~$1$ results in a significant change of the structure. 
  Similar structural change appears also for Series~$2$, but for lower $\bar T$: between $\bar T=0.30$ to $0.31$. As a result, the difference in the density is only a consequence of the shift of the structural transition, and the basic question is what is the reason of the shift. According to Test, the volume of  System~$1$ is large enough and the repulsive walls should not influence the transition. A much more probable reason for the nonphysical shift are the periodic boundary conditions applied for System~$2$. $L_x$ for System~$2$ amounts to only $28$, which is small considering the range of the potential ((\ref{uMD}), (\ref{uMD12})). To summarize, we can conclude that the most probable curve curse for the dense phase in equilibrium with the gas is that from Series~$1$ for $\bar T\le 0.32$, and Series~$2$ for $\bar T\ge 0.33$.

  \section{results}
  \label{results}
As already mentioned in sec.\ref{theoryA}, in the theory  we consider the interaction potential $V=u(6/\pi)^2$, with $u_{ii}=u=-u_{12}$ defined in Eq.(\ref{V(r)a}),  and in the simulations, $u_{ii}^{MD}$ and $u_{12}^{MD}$ are given in Eqs. (\ref{uMD}) and (\ref{uMD12}).  In both,  theory and simulations, we assume $A_Y=1.8$. The thermodynamic states are represented by the volume fraction, $\zeta=\frac{\pi}{6} N/V$, and temperature. As discussed in sec.~\ref{theoryB} and ~\ref{theoryC}, in the case of hard cores we should choose $T^*=kT/|\tilde V(k_0)|$. In the case of softened core studied in simulations, we present the results in terms of $\bar T=kT/\epsilon$. Our potential in Fourier representation takes the minimum $\tilde V(k_0)\approx -30.9\epsilon$ at $k_0\approx 1.33$.
\subsection{Theoretical results for the phase diagram}
\begin{figure}
\includegraphics[scale=0.35]{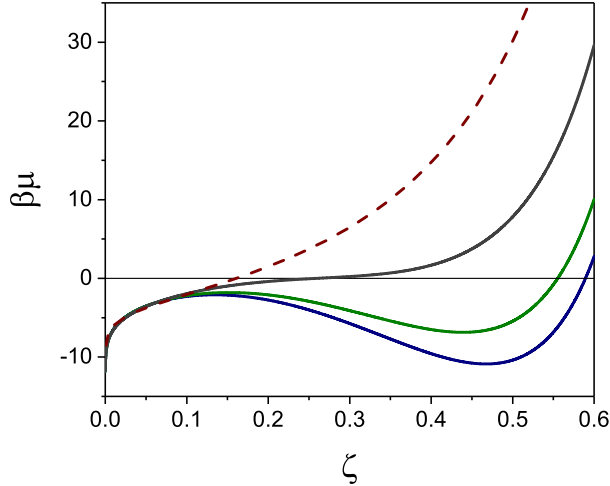}
\caption{The dimensionless chemical potential, $\beta\mu$, as a function of the volume fraction  in MF approximation. The dashed line corresponds to the disordered phase (Eq. (\ref{bmuMF})), and the solid lines correspond to the periodic phase (Eq.(\ref{mu})). From the top to the bottom solid line $T^*=0.11, 0.09, 0.086$, where temperature is in reduced units,
	$T^*=k_BT/|\tilde V(k_0)|$, 
 $k_0$ is the wavenumber corresponding to the minimum of the interaction potential $V=(6/\pi)^2u$ in Fourier representation 
 and $\zeta=\pi\rho/6$ is the dimensionless volume fraction, with the density $\rho=(N_1+N_2)/V$.
}
\label{muMF}
\end{figure}

Let us first discuss the theoretical results, and  focus on the MF approximation (sec.~\ref{theoryB}). The chemical potential isotherms for the disordered and the ordered phases are shown in Fig.~\ref{muMF}.  For the periodic phase ($\Phi>0$), we obtain 
from Eq.~(\ref{mu}) the chemical potential shape resembling simple fluids, with an unstable region for small $T^*$. This leads to a coexistence of two periodic phases with the same period, $2\pi/k_0\approx 4.7$, but different volume fractions.

\begin{figure}
\includegraphics[scale=0.5]{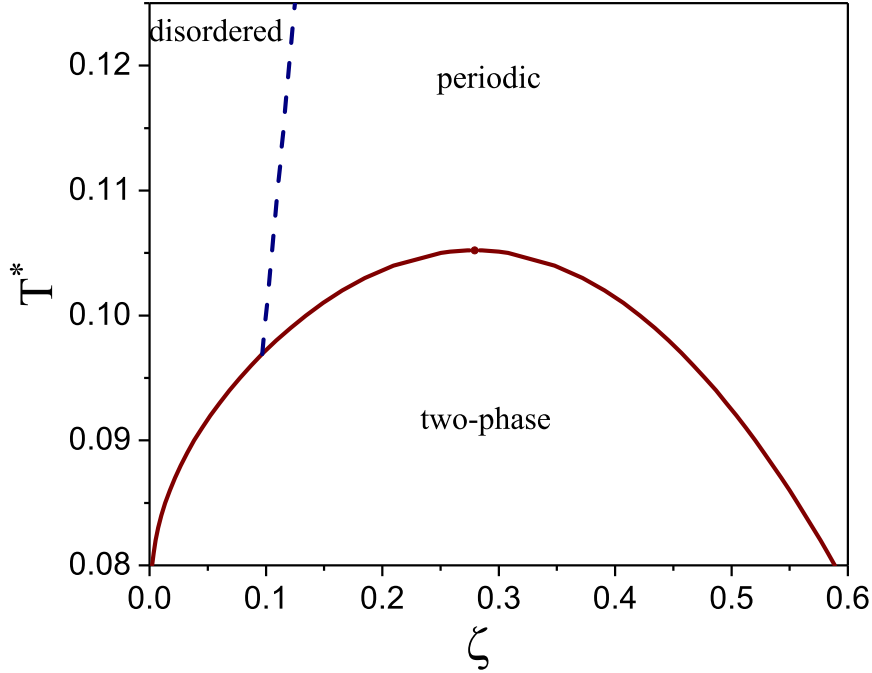}
\caption{The MF phase diagram of the model. Solid lines denote the first-order transitions, and the dashed line is the continuous transition between the disordered and periodic phases (the $\lambda$-line). The coexistence between the gas and the periodic phase occurs below the  temperature at the  intersection point  between the continuous and the first-order transitions, and above this temperature two  periodic phases, one  with low- and the other one with high density coexist. Temperature is in reduced units,
$T^*=k_BT/|\tilde V(k_0)|$, 
where $k_0$ is the wavenumber corresponding to the minimum of the interaction potential $V=(6/\pi)^2u$ in Fourier representation 
and $\zeta=\pi\rho/6$ is the dimensionless volume fraction, with the density $\rho=(N_1+N_2)/V$.}
\label{phd-MF}
\end{figure}

 To compute the phase diagram, we  compare $\beta\Omega_g$ (Eq.(\ref{Omg})) with $\beta\Omega_c$  (Eq.(\ref{Omaprox})) for fixed $T^*$, and with  $\bar\zeta(\beta\mu)$ obtained from Eqs.(\ref{bmuMF}) and (\ref{mu}). The intersection point between $\beta\Omega_g(\mu)$ and  $\beta\Omega_c(\mu)$ gives the coexistence of the two phases. 
The phase diagram in this MF approximation has the universal shape when the reduced temperature  $T^*$ is used, and is shown in Fig.~\ref{phd-MF}. 

\begin{figure}
\includegraphics[scale=0.35]{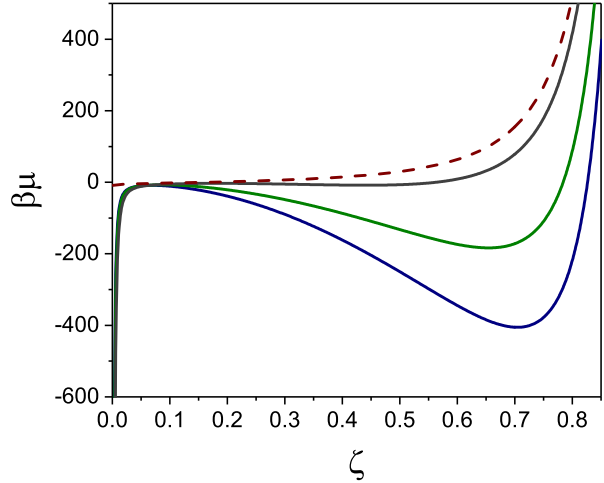}
\caption{The dimensionless chemical potential $\beta\mu$  as a function of the volume fraction $\zeta$ in the disordered phase. Dashed and solid lines represent the MF (Eq.~(\ref{bmuMF})) and the Brazovskii-type approximation (Eq.~(\ref{muf}) with $\Phi=\Psi=0$), respectively.  From the top to the bottom solid line $T^*=0.065, 0.0325, 0.025$.}
\label{mugfl}
\end{figure}

\begin{figure}
\includegraphics[scale=0.5]{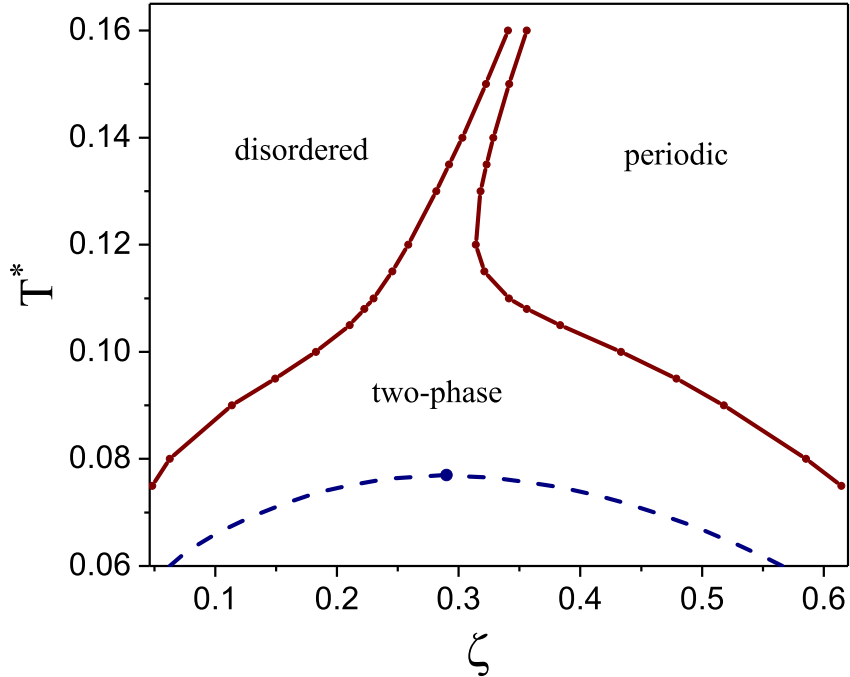}
\caption{Phase diagram in the  Brazovskii-type approximation. Solid lines are the first-order transitions between the disordered and ordered (periodic $c$) phases. The dashed line represents the metastable gas-liquid transition.  $\zeta$ is the average volume fraction of the particles, and $T^*=kT/|\tilde V(k_0)|$.
}
\label{phdB}
\end{figure}

Beyond MF, in the Brazovskii-type approximation, the chemical potential of the disordered phase is temperature-dependent (see Eq.~(\ref{muf})), and for low $T^*$ the unstable region of the volume fraction appears (see  Fig.~\ref{mugfl}). The presence of the instability leads to the gas-liquid separation that turns out to be metastable with respect to the phase transition between the disordered and ordered phases. The metastable gas-liquid transition with the asociated critical point is shown in Fig~\ref{phdB} as the dashed line. The order-disorder transition obtained by equating $\Omega(\mu)$ for the disordered and periodic phases is shown as the solid lines in Fig.~\ref{phdB}. According to our theory, the phase diagram in Fig.~\ref{phdB} corresponds to any system with hard-core particles interacting for $r>1$ with $V_{ii}=V=-V_{12}$, such that $\tilde V(k_0)^{''}/\tilde V(k_0)\approx -3$.  

At low $T^*$, the phase diagrams obtained in the  MF and in the Brazovskii-type approximations are similar, but at higher $T^*$ the two phase diagrams are qualitatively different. The continuous order-disorder transition  becomes fluctuation-induced first-order, and the dashed line in Fig.~\ref{phd-MF} is shifted to larger $\bar\zeta$ and transformed into the pair of lines enclosing the narrow two-phase region in Fig.~\ref{phdB}. The coexistence between the two periodic phases in MF is replaced by the order-disorder phase transition. Interestingly, the narrow two-phase region between the disordered and ordered phases broadens rapidly for $T^*$ decreasing from the value corresponding to the MF critical point of the  phase coexistence between the two ordered phases. We can identify the low-density ordered phase obtained in MF with the disordered phase in which the aggregates are self-assembled, but are not localized due to the presence of fluctuations.

Let us focus on the large-density branch of the phase coexistence, $T^*_c(\bar\zeta)$. For low $T^*$, the slope of this line is negative, and it changes sign for $\bar\zeta$ very close to its value at the MF critical point (see Figs.~\ref{phd-MF} and \ref{phdB}). This shape of the large-density branch of the phase coexistence has a strong effect on the structural evolution of a system with fixed number of particles for increasing $T^*$. For $\bar\zeta<0.3$, for example for $\bar\zeta=0.2$, the density of the periodic phase and its volume decrease, and the density and  volume of the coexisting gas increase when $T^*$ increases. Finally, when the low-density branch of the coexistence line is reached, the periodic phase disappears. 
When $\bar\zeta>0.3$, however, for example for $\bar\zeta=0.35$, the scenario is different. As long as the slope of the $T^*_c(\bar\zeta)$ line is negative, the system evolution upon heating resembles the scenario in the case of the gas-liquid coexistence. Namely, upon heating the density of the denser phase decreases, but its volume increases. When the high-density branch of the phase coexistence is reached, the coexisting fluid disappears and the concentration oscillations are present in the whole volume. The low-density periodic phase has many defects, mainly vacancies, in this temperature regime. Further heating leads to a nucleation of the disordered phase when the $T^*_c(\bar\zeta)$ line with the positive slope is reached (see Fig.~\ref{phdB} for $\bar\zeta=0.35$). The density of both the disordered and ordered phases increases with further increase of $T^*$, and the volume of the periodic phase decreases, until the disordered-phase branch of the coexistence line is met and the periodic phase disappears. In this temperature range, the disordering effect of the entropy of mixing plays a more important role, and the periodic phase becomes denser upon heating, in contrast to the temperature region corresponding to the negative slope of   $T^*_c(\bar\zeta)$, where the increase of $T^*$ leads to formation of vacancies and less dense packing of the particles. The amplitude $\Phi$ of the concentration oscillations decreases with increasing $T^*$, indicating less ordered states at higher $T^*$.
\subsection{Simulation results for the phase diagram}

\begin{figure}
\includegraphics[scale=0.5]{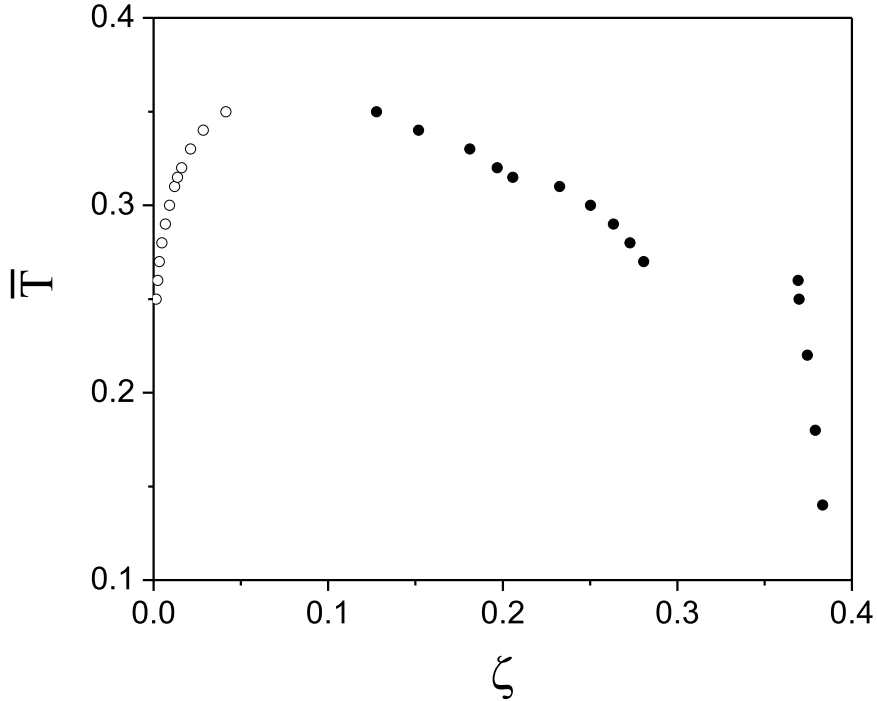}
\caption{The coexistence lines between the low- and high density phases, obtained in the MD simulations. $\zeta$ is the average volume fraction of the particles, and $\bar T=kT/\epsilon$. The structure of the dense phase at coexistence with the gas is shown in Figs.~\ref{snaphcp}-\ref{snap2}. For simulation details see sec.\ref{simulations}.
}
\label{pdMD}
\end{figure}

Let us focus on the results obtained in the MD simulations.
The coexistence lines between the low- and high density phases  obtained in simulations (see Table \ref{table}) are shown in Fig.~\ref{pdMD}. In the dense phase, we find oscillations of $c({\bf r})$ in one direction, in full agreement with assumptions of our theory. The period of the oscillations is very close to the theoretical prediction.  The negative slope of the high-density branch of the phase coexistence  agrees with the  low temperature-part of the theoretical phase diagram. The total volume fraction oscillations in the dense phase are not visible, again in agreement with the theory, where we get $\Psi\ll \Phi$ for the amplitudes of $\zeta$ and $c$. The volume fraction of the dense phase at the coexistence with the gas is lower than predicted in our theory. Note, however that the volume of the particle with the soft core is not uniquely defined, and the interaction potential takes a minimum for $r\approx 1.14$. Thus, the volume per particle is effectively larger than $\pi/6$ (length is in $a$-units), and it may lead to a larger volume fraction occupied by the particles than $\bar\zeta=\frac{\pi}{6} N/V$ shown in Fig.~\ref{pdMD}. 

\begin{figure}
\includegraphics[scale=0.55]{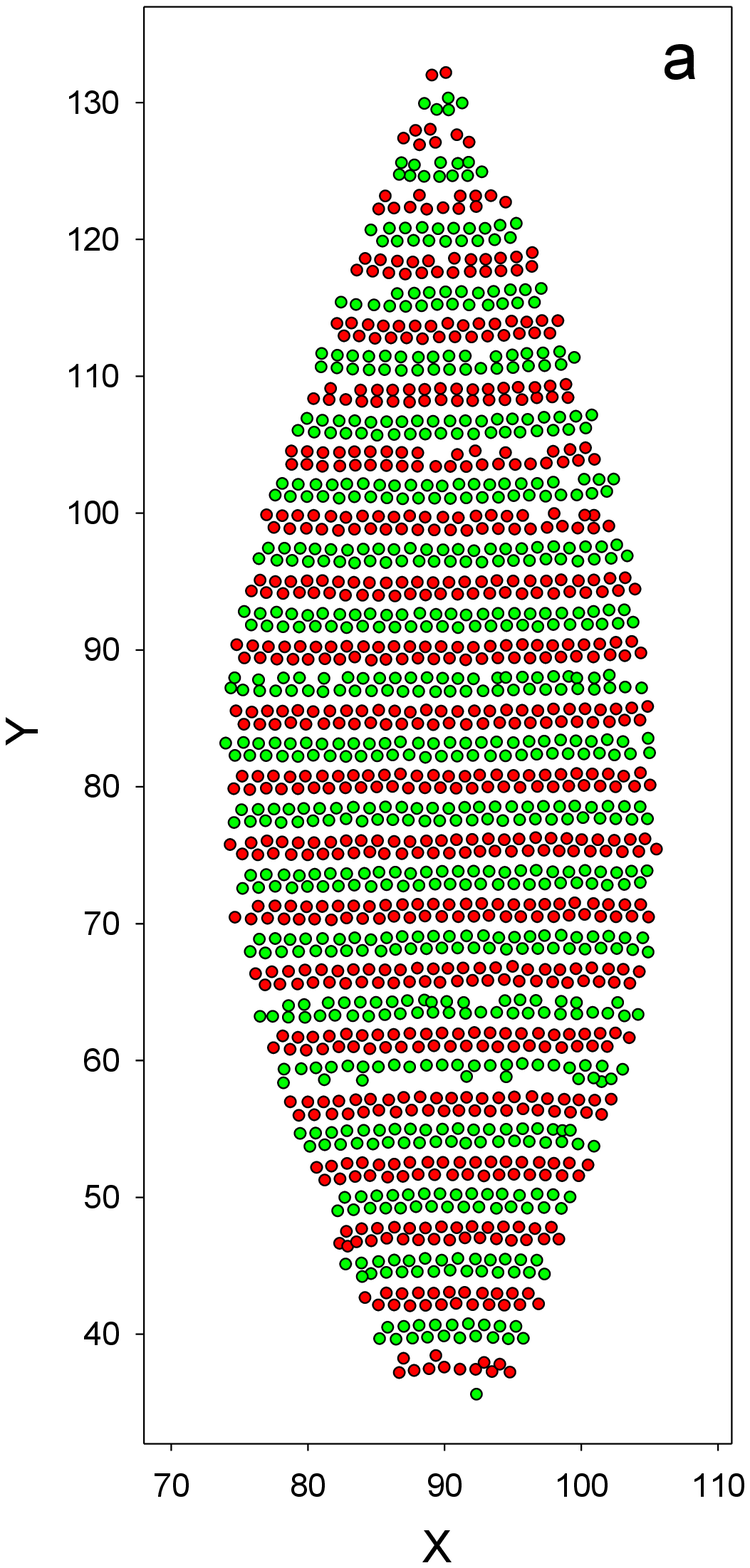}
\includegraphics[scale=0.5]{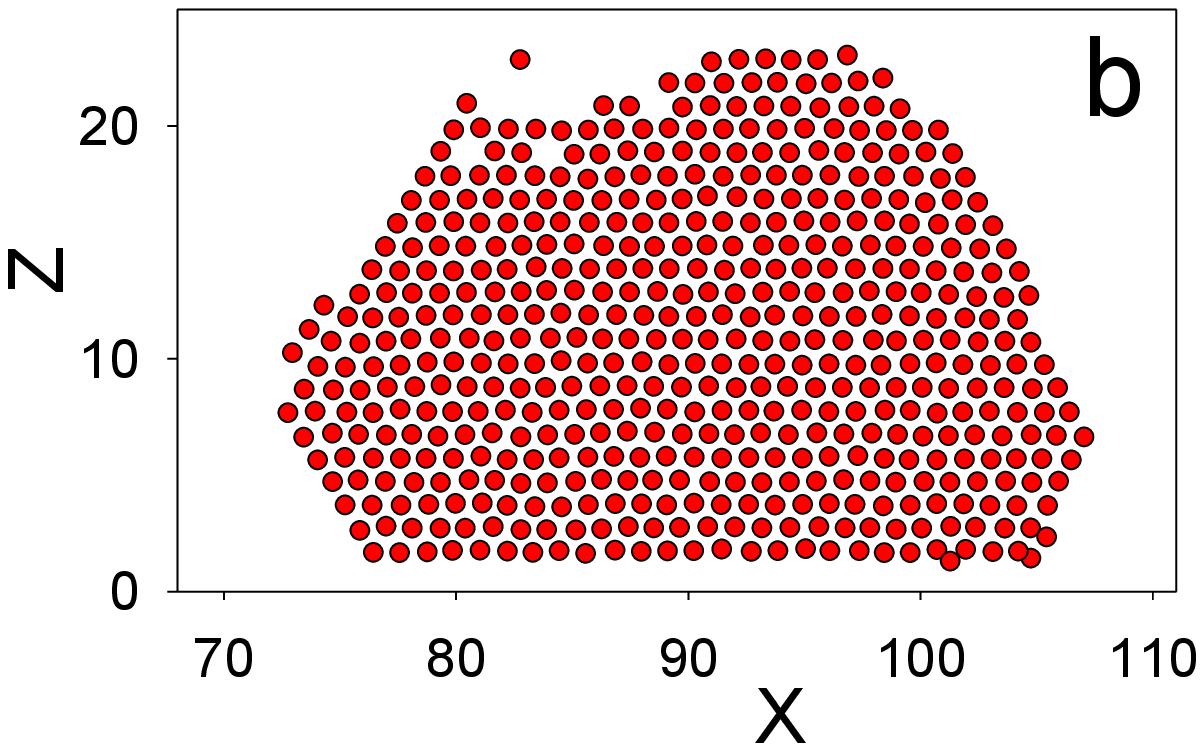}
\caption{The configuration obtained in the MD simulations at $\bar T=kT/\epsilon=0.16$.
Red and green circles with the diameter $\sigma=1.12$ (in $a$-units) represent particles of the first- and second component, respectively. A part of the simulation box  containing the monocrystal coexisting with a very dilute gas is shown.
(a) the projection of a layer of particles with $12<Z<13$ on the  $(X,Y)$ plane. (b) the projection  a layer of particles with  $76<Y<77$ on the  $(X,Z)$ plane. Note the perfect microsegregation of the particles in the bilayers and the hexagonal order in the  $(X,Z)$ plane. For simulation details see sec.\ref{simulations}.
}
\label{snaphcp}
\end{figure}

At very low $\bar T=kT/\epsilon$,  a very dilute gas coexists with  a  solid crystal, with the structure shown in the snapshots in Fig.~\ref{snaphcp}.  In Fig.~\ref{snaphcp}a, alternating bilayers of particles of the first species followed by bilayers of particles of  the second species can be seen. The shown bilayers are perpendicular to the direction of the oscillations of $c$. In Fig.~\ref{snaphcp}b,  one  layer belonging to the bilayer of particles of one species is shown. Note the hexagonal pattern formed by the particles. The ordering due to the packing effects of the particles
cannot be predicted by our theory in the considered approximation, and we do not expect agreement between our theory and simulations for large densities. 

\begin{figure}
\includegraphics[scale=0.5]{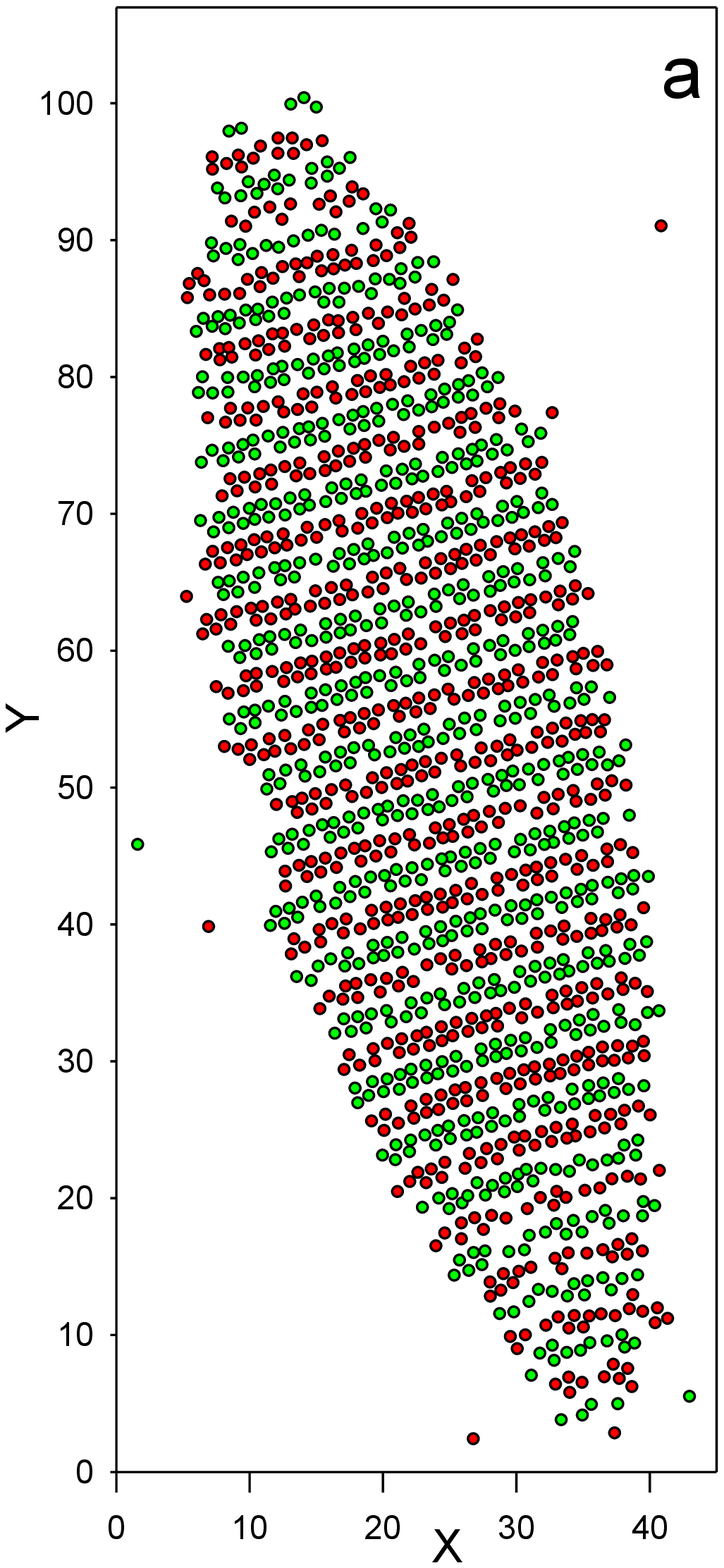}
\includegraphics[scale=0.5]{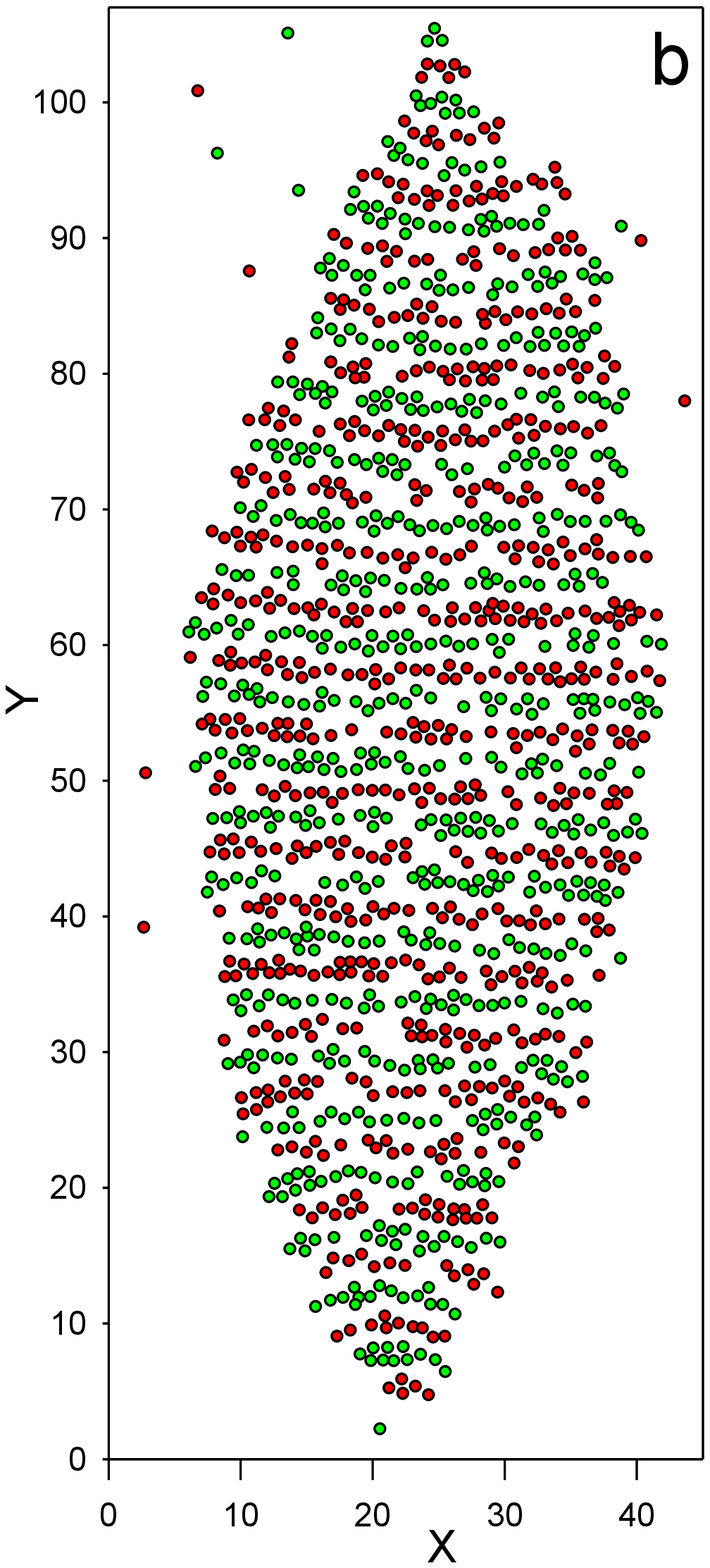}
\caption{ The  configurations obtained in the MD simulations for (a) $\bar T=kT/\epsilon=0.26$ and (b) $\bar T=kT/\epsilon=0.27$. Red and green circles with the diameter $\sigma=1.12$ (in $a$-units) represent particles of the first- and second component, respectively. The alternating layers rich in the first and in the second component are perpendicular to the shown planes. The particles for $\bar T\le 0.26$ form a crystal.   For $\bar T\ge 0.27$, the particle centers are disordered.   Note the smaller density and larger volume of the dense phase at $\bar T=0.27$, with the shape of the droplet swollen in directions perpendicular to the direction of oscillations of $c$. The discontinuity of the density at the coexistence with the gas is clearly seen in Fig.\ref{pdMD}. For simulation details see sec.\ref{simulations}.
}
\label{snap1}
\end{figure}

At $\bar T=0.27$, another periodic phase 
appears. This phase has a significantly lower density than the solid crystal, and has a structure of a soft- or liquid crystal, with alternating layers rich in particles of the first and the second component, but without positional order of the centers of the particles.
The structure of the crystal phase coexisting with the gas for $\bar T= 0.26$ is compared with the structure of the periodic phase coexisting with the gas for $\bar T= 0.27$  in Fig.~\ref{snap1}. 

\begin{figure}
\includegraphics[scale=0.5]{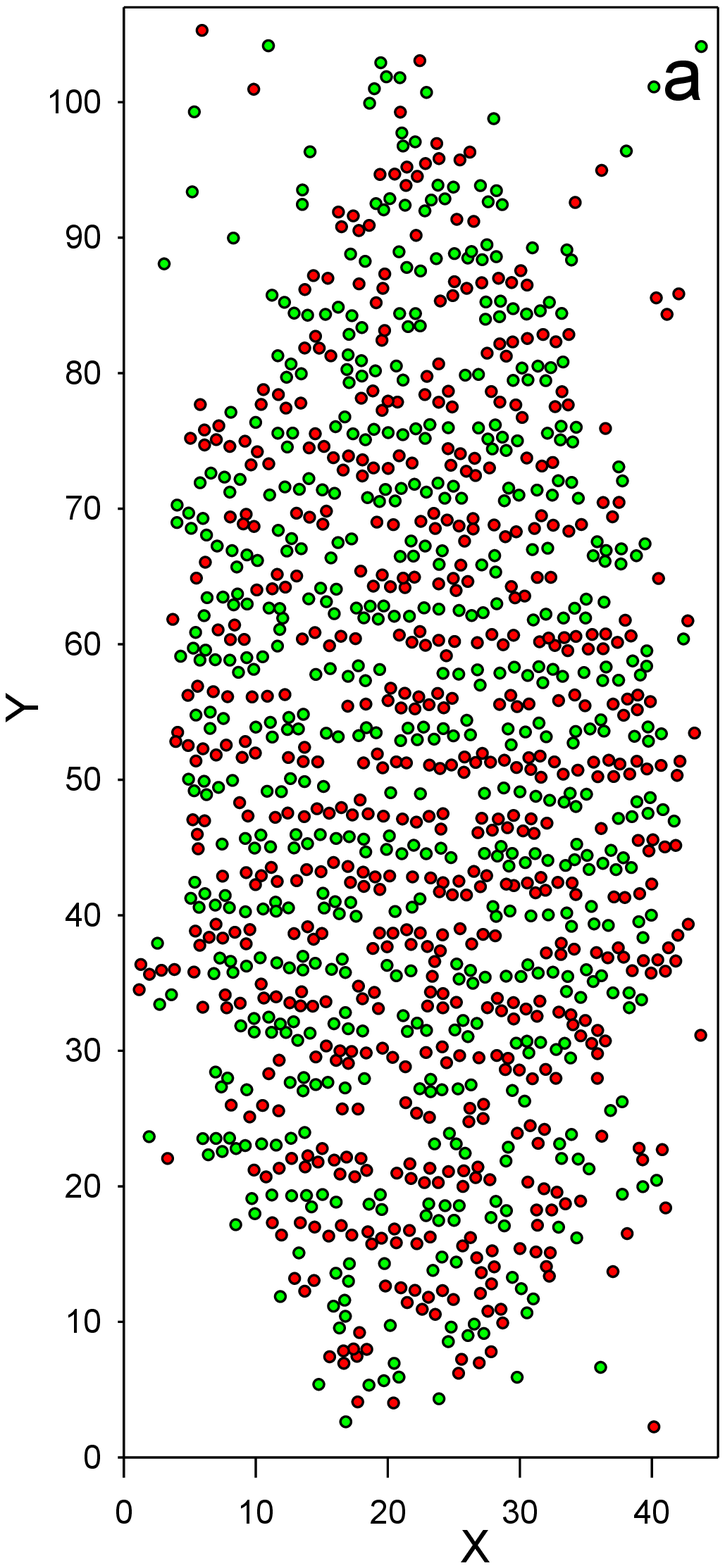}
\includegraphics[scale=0.5]{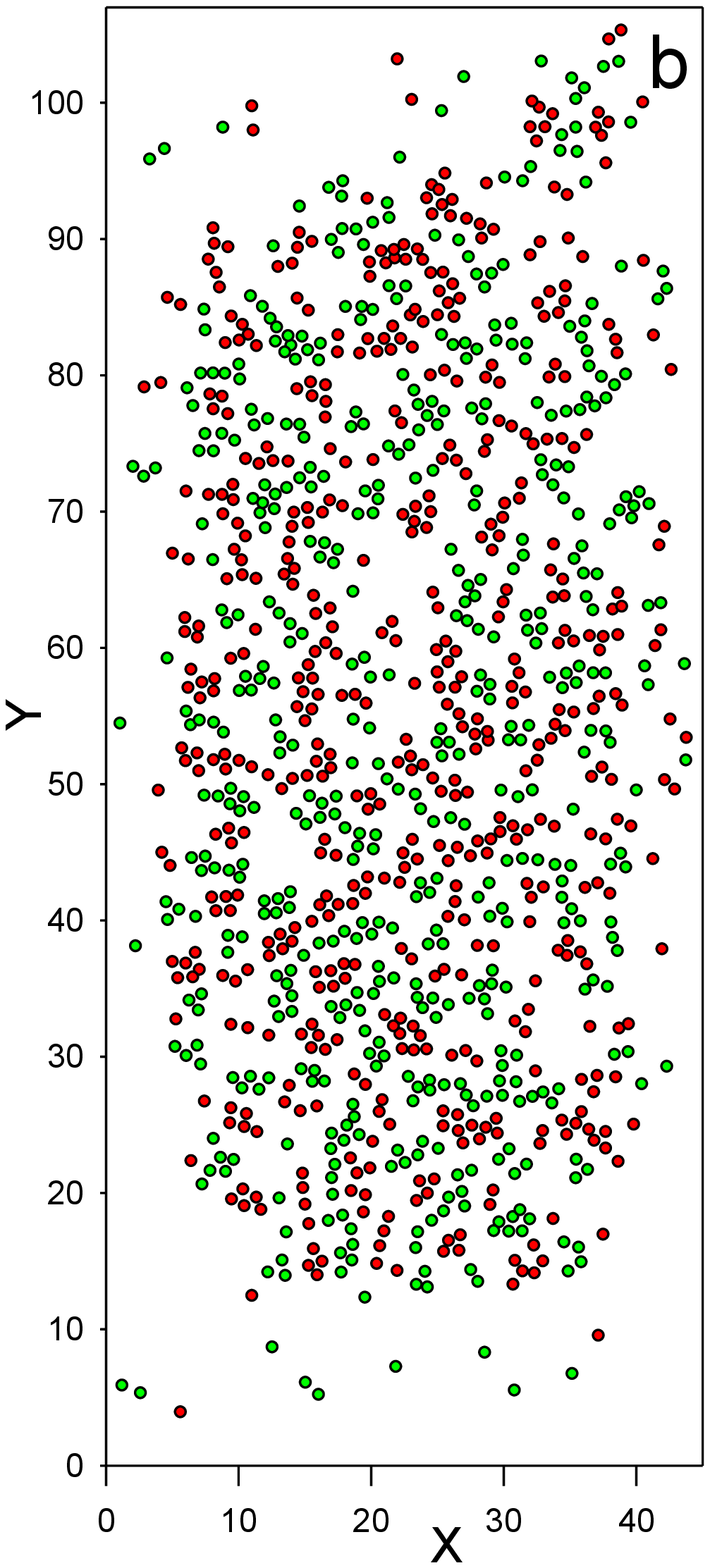}
\caption{ The  configurations obtained in the MD simulations for (a) $\bar T=kT/\epsilon=0.31$ and (b) $\bar T=kT/\epsilon=0.32$. Red and green circles with the diameter $\sigma=1.12$ (in $a$-units) represent particles of the first- and second component, respectively. The alternating layers rich in the first and in the second component, clearly seen at $\bar T=kT/\epsilon=0.31$, are perpendicular to the shown plane. For $\bar T=kT/\epsilon=0.32$ the microsegregation is still visible, but the snapshot appears less ordered. For simulation details see sec.\ref{simulations}.}
\label{snap2}
\end{figure}

Another structural change can be seen in Fig.~\ref{snap2}, where snapshots showing the droplet of the dense phase at coexistence with the gas are shown for $\bar T=0.31$ and $\bar T=0.32$. The structure of the dense phase at $\bar T=0.31$ is of the same type as at $\bar T=0.27$ (Fig.~\ref{snap1}b), and it differs noticeably from the structure at $\bar T=0.32$. In the latter case, the microsegregation of the components into aggregates of thickness $1$ or $2$ still takes place, but the concentration oscillations in one direction are no longer evident based on the visual inspection of the simulation snapshots.  More empty regions than at lower $\bar T$ occur, leading to a relatively large change of the density for temperature increasing by  $\bar T\sim 0.01$. 

\begin{figure}
\includegraphics[scale=0.5]{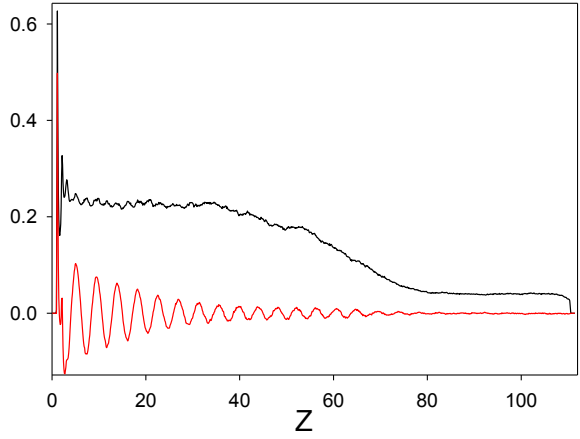}
\caption{
The total volume fraction (black line) and the concentration $c$ (red line), averaged over the $(X,Y)$ planes, at a distance $Z$ from the wall attracting the first and repulsing the second component for $\bar T=0.33$. All the remaining walls of the simulation box are repulsive (type (ii) BC). For simulation details see sec.\ref{simulations}.}
\label{g(z)}
\end{figure}

At this high $\bar T$ and low density, the droplet of the denser phase can undergo deformations, or even move as a whole during the simulations. In order to pin-point the droplet, we consider a wall attracting strongly the first component, and repulsing weakly the second component. The density and concentration profiles averaged over the planes $(X,Y)$ parallel to the wall are shown in Fig.~\ref{g(z)}. The ordering effect of the wall on the total density extends to short distances, while the concentration exhibits damped oscillations up to $Z\sim 40$, and for $Z>40$, the amplitude of the oscillations of $c$ remains almost constant, although quite small, in the denser phase. In the low density phase the oscillations of the concentration are not visible. The density profile is influenced by the shape of the droplet, since for increasing $Z$, the area in the $(X,Y)$ plane occupied by the dense phase decreases. We can conclude that the low-density phase is isotropic, but the weak periodic order of the denser phase  may still be present above $\bar  T=0.31$. Definite conclusions concerning the nature of the structural change at $\bar T\approx 0.31$, however, are not possible yet.  

Our simulation procedure does not allow for the determination of the coexistence lines for $\bar T>0.35$.

 As the number of particles in the simulations is fixed, we can observe the shape of the monocrystal or droplet coexisting with the gas, as shown in Figs.~\ref{snaphcp}-\ref{snap2}. The shape of the crystal or droplet for $\bar T\le 0.31$ indicates that the surface tension of the interface parallel to the microsegregated layers is  much larger than the surface
tension of the interface perpendicular to these layers. The droplet at $\bar T=0.32$ has a different shape, but it is still different from a sphere, expected for an isotropic liquid. 

\subsection{Structure at high temperature}
\begin{figure}
\includegraphics[scale=0.35]{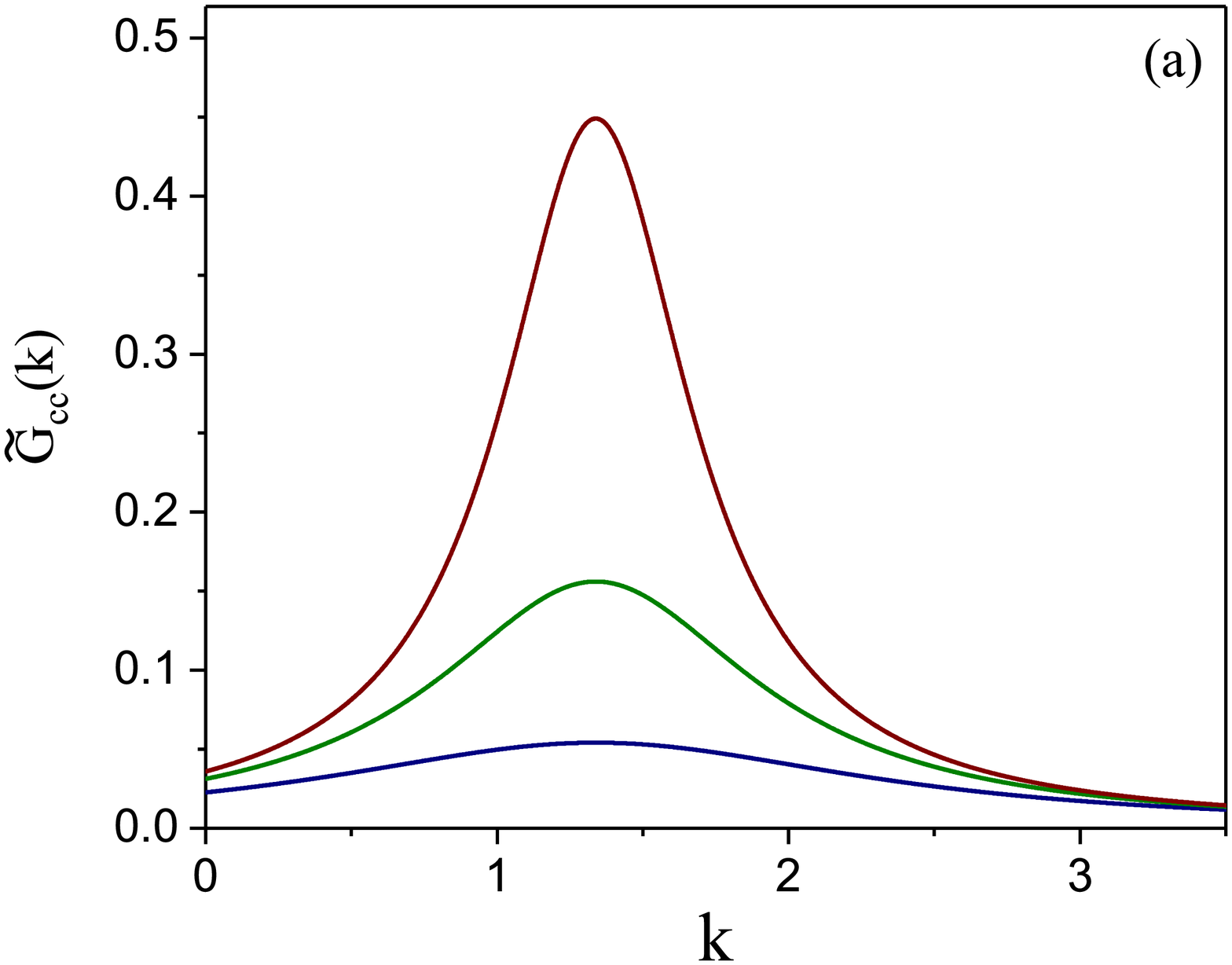}
\includegraphics[scale=0.35]{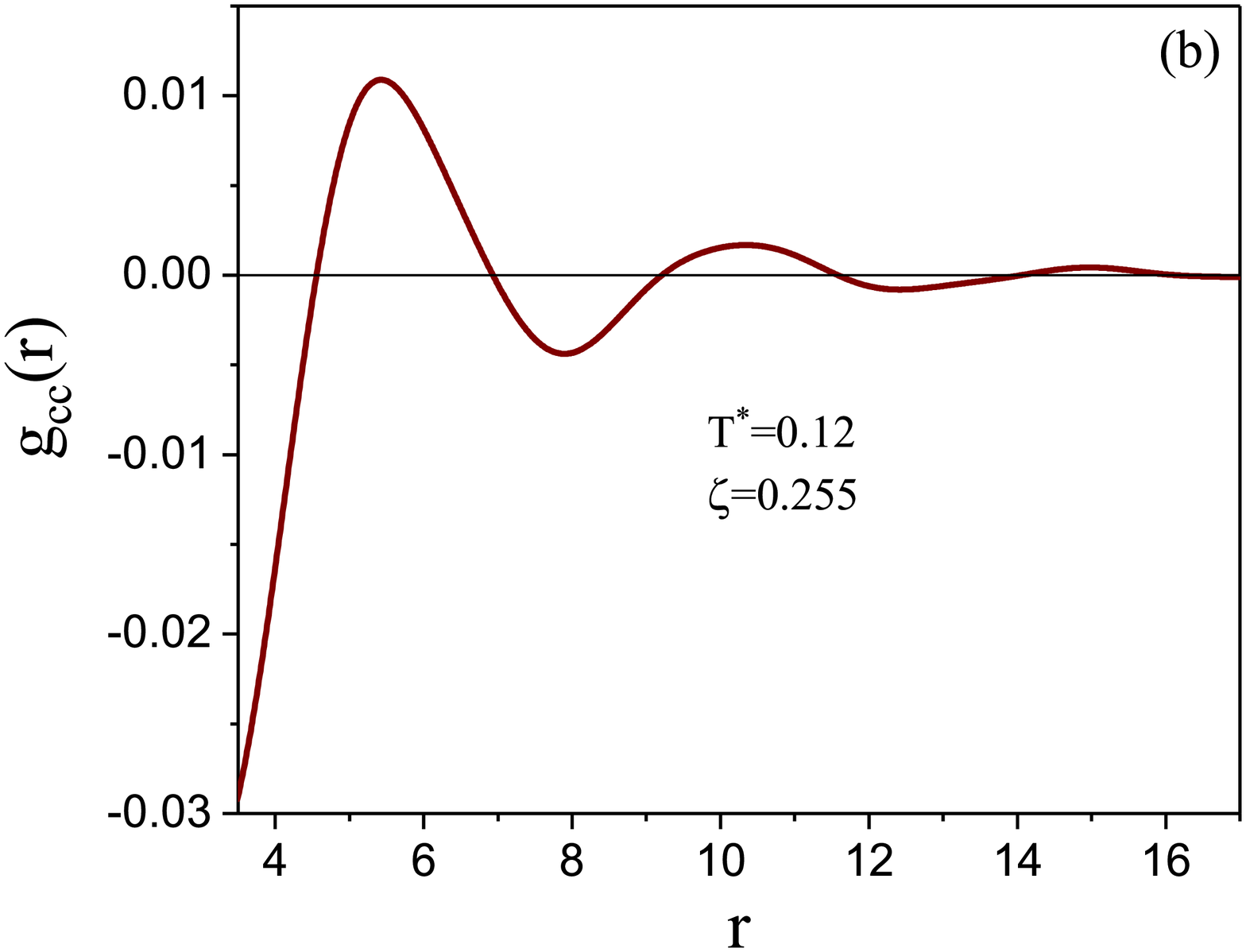}
\caption{The correlation function for the concentration obtained in the Brazovskii-type theory. (a) $\tilde G_{cc}(k)$ in Fourier representation for $T^*=0.11$. From the bottom to the top line, $\bar\zeta=0.1, 0.15, 0.2$. (b) $g_{cc}(r)$ (see Eq.~(\ref{gcc})) in real-space representation for $T^*=0.12$ and $\bar\zeta=0.255$.}
\label{Gkf}
\end{figure}

Let us focus on the structure of the disordered phase, and consider the correlation function.
In Fig.\ref{Gkf}a, we present $\tilde G_{cc}(k)$ 
 obtained in the Brazovskii-type approximation  for three values of the volume fraction and for $T^*=0.11$. We can clearly see the structure on the length scale $2\pi/k_0$ developing for increasing volume fraction, when the phase transition to the periodic phase is approached. For these thermodynamic states, $\tilde G_{cc}^{MF}(k)$ given in Eq.~(\ref{GMF}) does not exist, since in MF the disordered phase is unstable. 
In order to compare theoretical and simulation results, we plot the correlation function $g_{cc}$ related to the pair distribution functions $g_{ij}=G_{ij}\bar\zeta_i^{-1}\bar\zeta_j^{-1}+1$ according to the formula
\begin{equation}
\label{gcc}
g_{cc}(r)=(g_{11}(r)+g_{22}(r)-2g_{12}(r))\Bigg(\frac{\bar\zeta_1}{\bar\zeta}\Bigg)^2\Bigg(\frac{\bar\zeta_2}{\bar\zeta}\Bigg)^2.
\end{equation}
  $g_{cc}(r)$ is shown in Fig.\ref{Gkf}b  for $T^*=0.12$ and $\bar\zeta=0.255$. This thermodynamic state is close to the transition to the periodic phase, where the two-phase region becomes narrow. 

\begin{figure}
\includegraphics[scale=0.35]{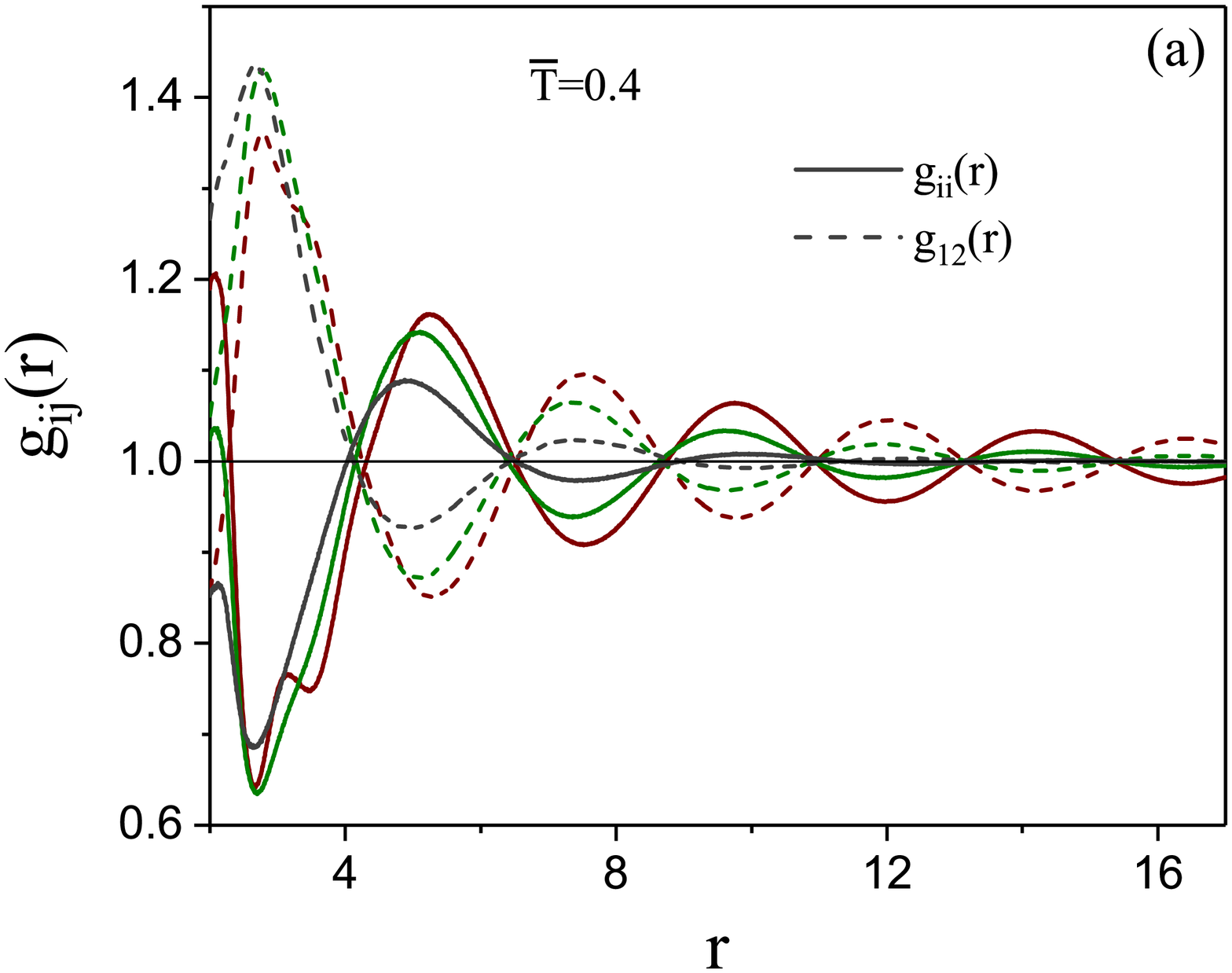}
\includegraphics[scale=0.35]{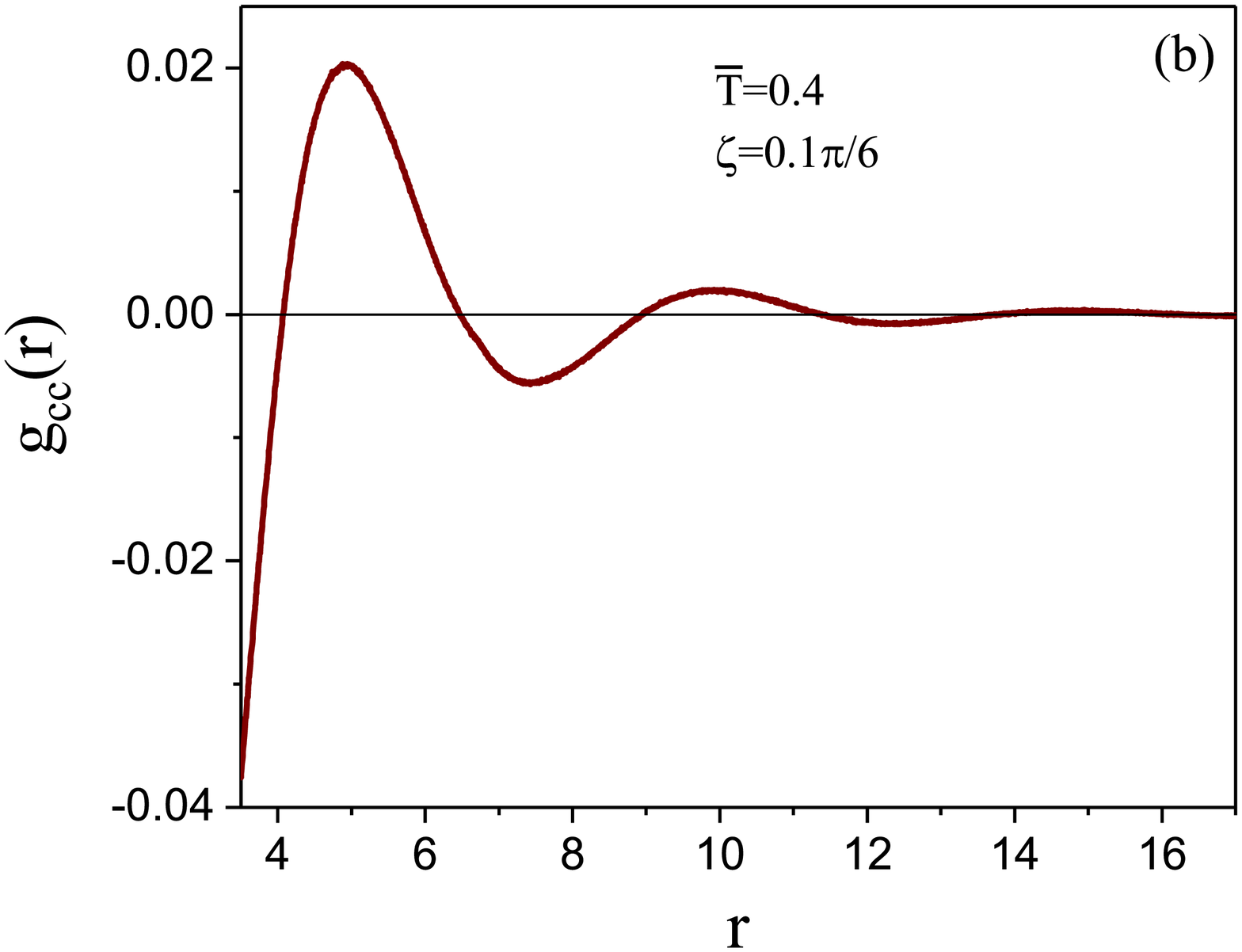}
\caption{MD simulation  results for (a) the correlation functions between like and different particles obtained in simulations for $\rho=N/V= 0.4,0.2, 0.1$ and  $\bar T=kT/\epsilon=0.4$, with periodic BC (type (iv)). Representative configurations corresponding to $\rho=0.1$ and $\rho=0.4$ are shown in Fig.~\ref{snap3}. (b) the correlation function for concentration, $g_{cc}$ (see Eq.~ (\ref{gcc})), 
 at $\zeta=0.1\pi/6$ and $\bar T=0.4$. For simulation details see sec.\ref{simulations}.}
\label{gijMD}
\end{figure}

The pair distribution functions obtained in MD for three values of density at $\bar T =kT/\epsilon=0.4$ are shown in Fig.~\ref{gijMD}. We can see  increasing correlations when the volume fraction increases. For comparison with the theoretical results, we plot  in Fig.~\ref{gijMD}b  $g_{cc}$
 at $\zeta=0.1\pi/6$ and $\bar T=0.4$. This thermodynamic state corresponds to the low-density phase not far from the phase transition. Note the very similar period and decay length close to the transition to the ordered phase in theory and simulations (Fig.~\ref{Gkf}b and Fig.~\ref{gijMD}b). For comparison of the thermodynamic states shown in Fig.~\ref{Gkf}b and Fig.~\ref{gijMD}b, see Figs.\ref{phd-MF} and \ref{phdB}.

\begin{figure}
\includegraphics[scale=0.35]{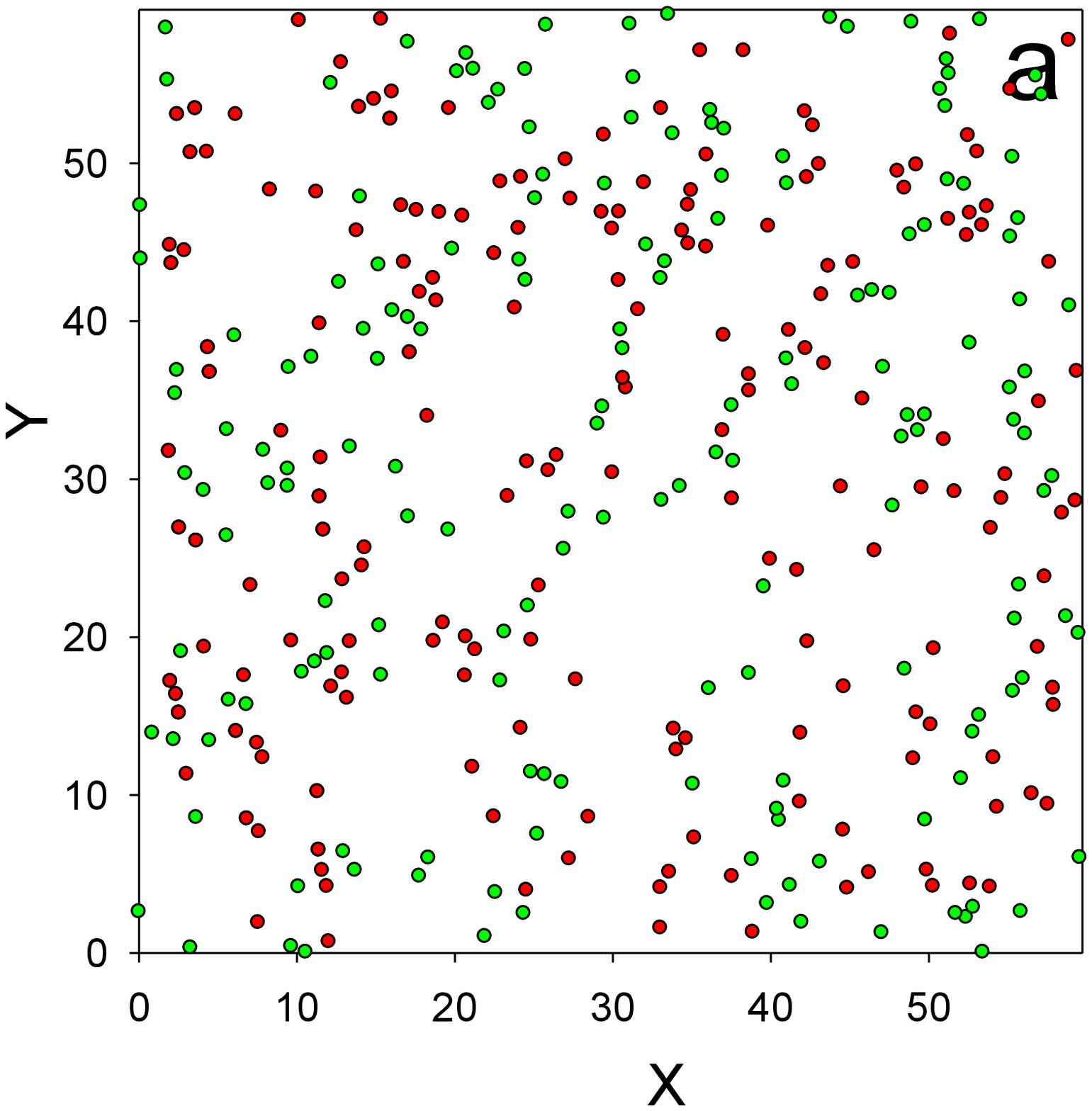}	
\includegraphics[scale=0.35]{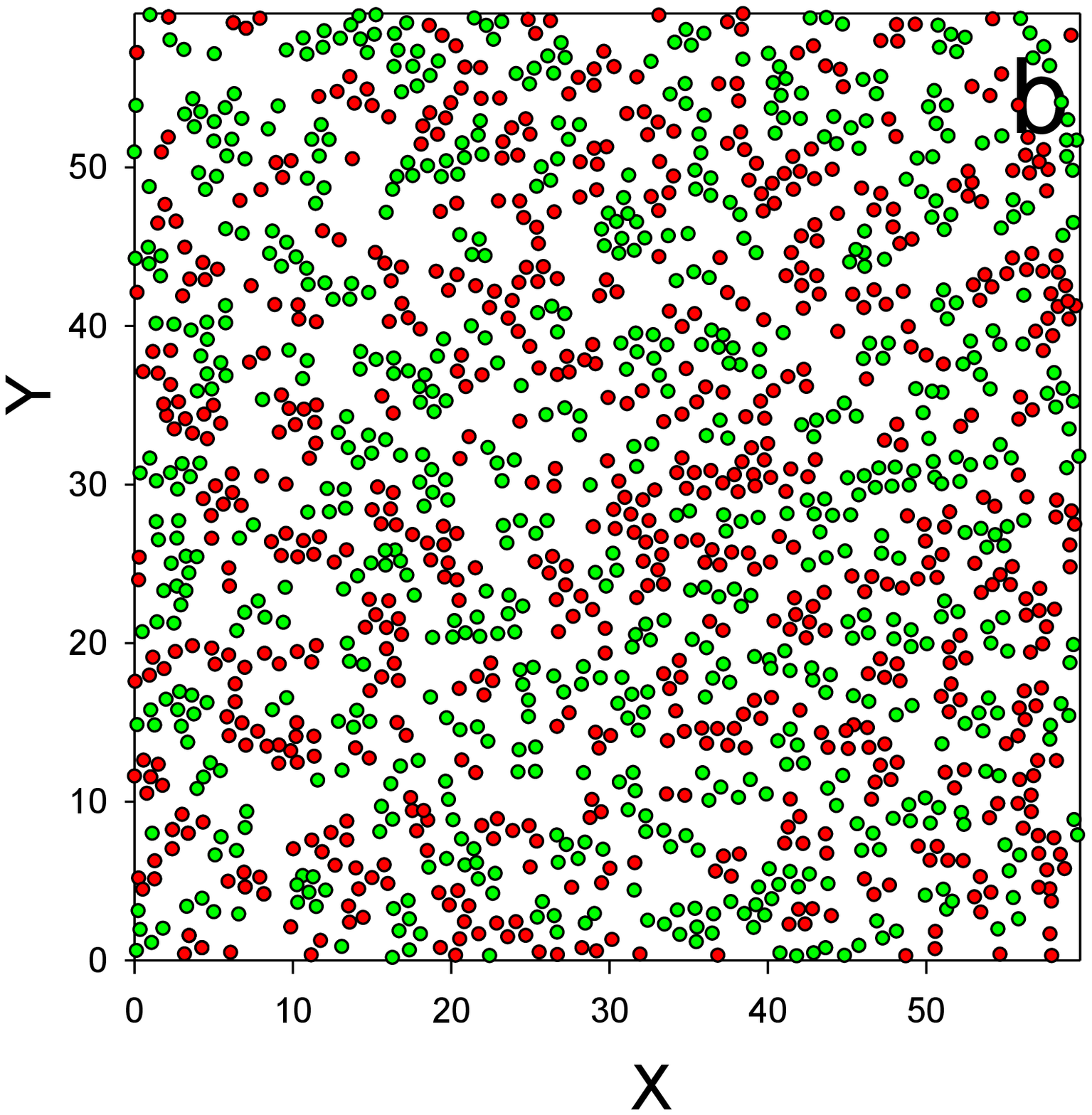}
\caption{ A projection of a layer of thickness $1$ on the $(X,Y)$ plane showing the configuration obtained in the MD simulations for  $\bar T=kT/\epsilon=0.4$, with periodic BC (type (iv)). (a)  $N/V=0.1$ and (b) $N/V=0.4$. The corresponding pair distribution functions are presented in Fig.\ref{gijMD}. }
\label{snap3}
\end{figure}


  \section{discussion and summary}
  
  We have studied phase behavior and structure in a binary mixture of particles with competing interactions, assuming hard cores of the particles in the theory, and core-softened particles in the MD simulations. The assumed interactions favor close neighbors of the same kind, but  at larger distances the presence of different particles is favorable. We have obtained good agreement between the theory developed in this work and MD simulations. In our theory, local fluctuations of the concentration are taken into account for the first time in the  formalism that allows to determine the phase diagram in microsegregating mixtures.
  
 At low $T$, the simulations show formation of the crystal phase with perfectly separated particles in alternating bilayers composed of the first and the second component.  We did not obtain the crystal in our mesoscopic theory, because we assumed weak order that is present at higher $T$. 
 
 At $\bar T=0.27$, the liquid-crystalline  phase with periodic concentration appears in the simulations, and this phase coexists with the gas at  $\bar T\ge 0.27$. In this periodic phase, alternating layers rich in the first and the second component are formed, but the crystalline order of the centers of mass of the particles is absent. Further heating leads to lower density and smaller degree of order of the periodic phase at the coexistence with the disordered phase. 
   At these intermediate temperatures, the theoretical and simulation results are in good agreement.
  
The phase diagram  at large temperatures cannot be reliably determined in our simulations. For this reason, the theoretical predictions for the high-temperature part of the phase diagram are not supported by simulations yet. The periodic structure with weak order, expected at high $T$, is characterized by the ensemble-averaged  concentration that is periodic in space with a small amplitude. Small amplitude  means a large number of different defects in instantaneous states due to thermally induced fluctuations. More subtle simulation methods are necessary for detection of the weak order and determination of its nature.
  
  On the quantitative level, the volume fractions and temperature at the phase diagrams obtained in the theory and in the simulations are significantly different. This is partially because of the approximate nature of the theory, and partially because of the softer core in simulations. The diameter of the particle core in simulations is not uniquely defined, and the potential takes the minimum for a distance noticeably larger than the diameter of the hard core considered in the theory. Thus,
  we should remember that the actual volume occupied by the particles is larger than shown in Fig.\ref{pdMD}. The difference in temperature corresponding to the phase coexistence of the gas and the dense liquid crystal is as large as one order of magnitude. For example, $\bar T=kT/\epsilon =0.31$ in simulations corresponds to $T^*\approx kT/(31\epsilon)\approx 0.01$ in the units used in our theory. The corresponding thermodynamic state in Fig.\ref{phdB}, however, occurs for $T^*\sim 0.1$. This discrepancy is related mainly to the minimum of the interaction potential that for hard cores is about $10$ times deeper than in the case of the softer core (compare Fig.\ref{figVk}a and \ref{figVkMD}a). Because of that, the thermal energy $kT$ becomes comparable with the depth of the attractive well for $T$ that in the case of hard cores is about $10$ times larger than in the case of the softer core. This means that we need $\sim 10$ times larger temperature at the order-disorder transition when the softer core is replaced by the hard one. Note that when the shape of the interactions at short distances is taken into account, our theoretical and simulation predictions are in good agreement. Also, the correlation function calculated for $T^*=0.12$ in the theory and for $\bar T=0.4$ in the simulations, correspond to very similar temperature, if $kT$ is in units of the minimum of the potential between like particles.
  
The  strong effect of the softness (or hardness) of the particle core on the temperature at the order-disorder phase transition should be taken into account in experiments searching for spontaneously formed ordered patterns on the nanometer or micrometer length scale.

Experimental observation of ordered patterns in binary colloidal mixtures concerns colloidal crystals formed by oppositely charged particles~\cite{blaaderen:05:0} with cubic symmetry. The same crystals with rich variety of unit cells were found in computer simulations~\cite{blaaderen:05:0}. Another example of pattern formation  in binary mixtures concerns colloidal gels~\cite{appel:15:0}.
As far as we know, however, colloidal crystals and liquid crystals with the components microsegragated into alternating planar layers have not been observed yet. This new type of ordered phases may be found for example in a mixture  with the mermaid potential between like particles (attractive head repulsive tail), and the peacock potential between different ones (repulsive head attractive tail).
 From the theory developed here it follows that the topology of the phase diagram is common for many systems with this type of interactions. The strength of the interactions should be carefully designed to obtain the ordered phases at room temperature. Also the thickness of the microsegragated layers can be controlled by tuning the interaction potential. 
 
 \section{ Acknowledgments}
This project has received funding from the European Union Horizon 2020 research and innovation under the Marie
Sk\l{}odowska-Curie grant agreement No 734276 (CONIN). An additional support in the years 2017--2020  has been granted  for the CONIN project by the Polish Ministry of Science and Higher Education. 


 \end{document}